\documentclass[showpacs,preprintnumbers,amsmath,amssymb]{revtex4}
 \usepackage{amssymb,amsmath}
 \usepackage{graphicx,epsfig}
 \usepackage{dcolumn}
 \usepackage{bm}

\begin{document}

 \title{Quantum motion of a neutron in a
 wave-guide in the gravitational field.}
 \author{A.Yu. Voronin}
 \affiliation {P.N. Lebedev Physical Institute, 53 Leninsky
 prospekt,119991, Moscow, Russia}
 \author{H. Abele}
 \affiliation {Physikalisches Institut der Universit\"{a}t Heidelberg\\
Philosophenweg 12\\
69120 Heidelberg, Germany }
 \author{ S. Bae{\ss}ler}
 \affiliation {Institut of physics,
University of Mainz, 55099 Mainz, Germany}
 \author{V.V. Nesvizhevsky}
 \author{A.K. Petukhov}
 \affiliation{Institut Laue-Langevin (ILL), 6 rue Jules Horowitz,
 F-38042, Grenoble, France}
 \author{K.V. Protasov}
 \affiliation { Laboratoire de Physique Subatomique et de Cosmologie
 (LPSC), IN2P3-CNRS, UJFG, 53, Avenue des Martyrs, F-38026, Grenoble,
 France}
 \author{A. Westphal}
 \affiliation {ISAS-SISSA and INFN, Via Beirut 2-4, I-34014 Trieste, Italy}
 \begin{abstract}
 We study theoretically the quantum motion of a neutron in a
 horizontal wave-guide in the gravitational field of the Earth. The
  wave-guide in question is equipped with a mirror below and a
 rough surface absorber above. We show that such a system acts as a quantum
 filter, i.e. it effectively absorbs quantum states with sufficiently high
transversal energy but transmits low-energy states.
 The states transmitted are mainly determined by the potential
 well formed by the gravitational field of the Earth and the mirror. The
formalism developed for quantum motion in an absorbing wave-guide is
 applied to the description of the recent experiment on the
 observation of the quantum states of neutrons in the Earth's
 gravitational field.
 \end{abstract}
 \maketitle
 \section{Introduction}
 Although the solution of the problem of the quantization of particle
motion in a well formed by a linear potential and ideal mirror has
been known for a long time
\cite{Gold,Haar,Flugge,Langhoff,Gibbs,Sakurai} the experimental
observation of such a phenomenon in the case of a gravitational
field is an extremely challenging task.

 The electric neutrality of neutrons \cite{Luschikov1,Luschikov2,Shull,Gahler,Bau} is an
advantage for this kind of research. Thus, in earlier experiments
the use of cold neutrons has allowed the gravitationally induced
phase-shift of neutrons to be measured
\cite{Colella,Rauch,Stau,Lit,Zouw}.

\begin{figure}
  \centering
 \includegraphics[width=115mm]{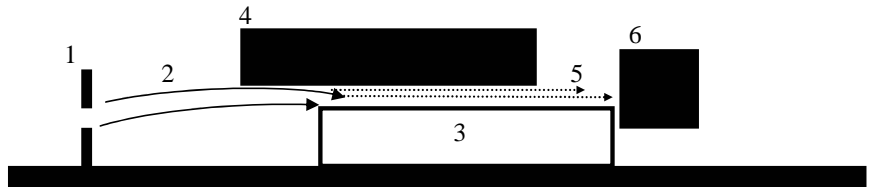}
  \caption{Schematic view of the experiment. From left to
the right: the vertical bold lines indicate the upper and lower
plates of the input collimator (1); the solid arrows correspond to
classical neutron trajectories (2) between the input collimator and
the entry slit between a mirror (3, empty rectangle below) and a
scatterer (4, black rectangle above). The dotted horizontal arrows
illustrate the quantum motion of neutrons above a mirror (5), and
the black box represents the neutron detector (6).}\label{FigInst}
 \end{figure}

 The direct observation of the lowest quantum states of neutrons in the
 Earth's gravitational field above a mirror has recently become
 possible.
The experiment consists of the measurement of the neutron flux
through a slit between a mirror and an absorber (scatterer) as a
function of the slit size. Slit size could be finely adjusted and
precisely measured. The neutron flux in front of the experimental
installation (in Fig.\ref{FigInst} on the left) is uniform over
height and isotropic over angle.  A low-background detector measures
the neutron flux at the  exit (in Fig.\ref{FigInst} on the right).
The main aim of this experiment was to demonstrate, for the first
time, the existence of the quantum states of matter in a
gravitational field. The detailed description of the
 experiment and a discussion of its reliability and precision can be found
 in refs. \cite{Netal00,Netal1,Netal2,Netal3,Netal4,
 Sweden,Answer,Bowles,Schw,Nanopart,NesvUspekh}.

 The gravitationally bound quantum states of neutrons and the related
experimental techniques provide a unique tool for a broad range of
investigations in the fundamental physics of particles and fields.
These include the Equivalence Principle tests in the quantum domain
as well as short-range fundamental forces studies
\cite{Murayama,Bertolami,Bert1,EQ1,EQ2,EQ3,Abele,NP1,NP2} and the
study of
 the foundations of quantum mechanics \cite{Mavromatos,Rob}. The experiment
on neutron gravitational quantum states stimulated progress in
surface studies (see, for instance,\cite{Techn1,Techn2}). A short
overview of the applications can be found in\cite{Apllications}.

These studies require clear understanding of the quantum mechanical
problem of neutron passage through an absorbing wave-guide in the
presence of gravitational field. Here we develop a theoretical model
of neutron quantum motion in such a wave-guide.

 In Chapter II we summarize the main known facts about a solution of the quantum-mechanical problem for
a particle in the potential well formed by a linear potential and an
ideal horizontal mirror. In Chapter III we discuss the main
principles of observation of neutron quantum states using the
absorbing wave-guide. We show that such a wave-guide turns out to be
a quantum filter, which  absorbs states with high transversal energy
and  transmits low-energy states. These transmitted states are
mainly determined by the potential well formed by the gravitational
field and the mirror.

 The latter condition is a specific feature of our problem, which, to our
knowledge, has not been explicitly considered in the literature
(see, for instance, refs. \cite{RS1,RS2,RS3,RS4, RS5} and the
references therein, devoted to the theory of the interaction of
waves with rough surfaces). Chapter IV is devoted to the passage of
neutrons through the wave-guide with a flat neutron absorber, as
proposed in \cite{Luschikov1,Luschikov2}, and Chapter V to their
passage with a rough absorber, as proposed in \cite{Netal00}. We
examine several models for the mechanism of neutron loss as a result
of their interaction with an absorber and discuss the limits of
their validity.

 The final chapter summarizes the conclusions. The results obtained are
rather general in character and can be applied to different physical
problems, involving the transmission of quantum particles through
absorbing wave-guides.

 \section{Quantum bouncing above mirror in the gravitational field}
 Although the results of this section can be found in the handbooks
\cite{Gold,Haar,Flugge} it is convenient to have them at our
disposal here. We start with a well-known problem of a particle
bouncing in the gravitational field above a perfect reflecting
mirror. In the following we consider $\hbar=1$. The characteristic
for this problem energy scale $\varepsilon_0$ and length scale $l_0$
are:
 \begin{eqnarray}
 \varepsilon_0&=&\sqrt[3]{ m g^2/2}\label{escale}\\
 l_0&=&\sqrt[3]{ 1/(2m^{2}g)}\label{lscale}
 \end{eqnarray}
 here $m$ represents the particle mass, and $g$ free fall acceleration. In
 the case of neutrons, which will interest us below, these quantities are:
 \[\varepsilon_0=0.602 \mbox{ peV, } l_0=5.871 \mbox{ } \mu m\]

 The Schr\"{o}dinger equation, which governs the wave-function of the
 neutron, confined between mirror and gravitational field is:
 \[-\frac{d^2\varphi_n(\xi)}{d\xi^2}+\xi \varphi(\xi)=\lambda_n
 \varphi(\xi) \]
  where dimensionless variable $\xi$ is connected with the distance
  variable $z$ via $\xi=z/l_0$, while quantum number
  $\lambda_n$ determines the energy values
  $\varepsilon_n=\varepsilon_0 \lambda_n$.
  The obvious boundary conditions are:
 \begin{eqnarray}
 \varphi_n(0)&=& 0 \label{boundzero}\\
 \varphi_n(\infty)&=&0\label{boundinf}
 \end{eqnarray}

 The wave-functions which satisfy the equations above are
 known to be:
 \begin{equation}\label{Ai}
 \varphi_n(\xi)\sim \mathop{\rm Ai}(\xi-\lambda_n)
 \end{equation}
 here $\mathop{\rm Ai}$ is the Airy function \cite{AbSt}.
 Substitution of (\ref{Ai}) into (\ref{boundzero}) gives the equation
 for the eigenvalues $\lambda_n$:
 \begin{equation}\label{lambda0}
 \mathop{\rm Ai }(-\lambda_n)=0
 \end{equation}

 The semiclassical (WKB) expression for the eigenvalues is:
 \begin{equation}\label{WKB}
 \lambda_n^{WKB}=\left(\frac{3\pi}{4}(2n-1/2)\right)^{2/3}
 \end{equation}
 This approximation gives the eigenvalues with accuracy to a few percent
 even for the lowest $n$.

  The asymptotic behavior of the gravitational states' wave-functions in the
  classically forbidden region $\xi\gg \lambda_n$ is characterized by very
fast decay:
  \begin{equation}\label{AssAi}
 \mathop{\rm Ai}(\xi-\lambda_n)\sim\exp(-2/3 (\xi-\lambda_n)^{3/2})
 \end{equation}
 The fast decay of the wave-functions under the gravitational
 barrier allows us to introduce a well-defined characteristic distance
 $H_n=l_0\lambda_n$ of a given state, which corresponds to the
 classical turning point $H_n=E_n/(Mg)$ of a bouncing particle with
 a given energy. Thus the quantization of energy $E_n$ is reflected in spatial distribution of the neutron density
 in the above-mentioned states (hereafter referred to as
 gravitational states). The scanning of
 this "quantized" spatial distribution of neutron density can be used
 to observe neutron quantum motion experimentally in the gravitational field.

 In Table \ref{Table1} we present the first seven eigenvalues
 $\lambda_n$, their WKB approximation $\lambda_n^{WKB}$ together with
 the corresponding energy values $E_n=\varepsilon_0 \lambda_n$
 and classical turning points $H_n=l_0\lambda_n$.
 \begin{table}
 \centering
 \begin{tabular}{|c|l|l|l|l|}
  \hline
  $n$ & $\lambda_n$ & $\lambda_n^{WKB}$ & $E_n$, peV & $H_n$, $\mu m$\\
  \hline
  1 & 2.338 & 2.320 & 1.407 &13.726 \\
  2 & 4.088 & 4.082 & 2.461 & 24.001\\
  3 & 5.521 & 5.517 & 3.324 & 32.414\\
  4 & 6.787 & 6.784 & 4.086 & 39.846\\
  5 & 7.944 & 7.942 & 4.782 & 46.639 \\
  6 & 9.023 & 9.021 & 5.431 &52.974\\
  7 & 10.040& 10.039& 6.044 &58.945 \\
  \hline
 \end{tabular}
 \caption{Eigenvalues, gravitational energies and classical turning
 points of neutrons in the earth's gravitational field above a mirror
 } \label{Table1}
 \end{table}


 \section{The principle for observation of the quantum gravitational
 states}

  Here we discuss only the principle of the experimental observation of
  neutron gravitational states based on the concept of neutron tunneling
 through the gravitational barrier, which separates the classically allowed
 region  and the absorber position \cite{Netal3,Netal4}.

  A flux of neutrons with horizontal velocity $V$ (from $4$ to $10$ m/s) was
 driven through a slit of variable height between a perfect horizontal
 mirror and a highly efficient absorber placed parallel to
 the mirror. The length $L$ of the wave-guide (which varied in
 different measurements from $L=10$ to $L=20$ cm) determined the
 neutron passage time $\tau^{pass}=L/V\simeq 2\mbox{ }10^{-2}$ s. It
 was found that when the slit height $H$ was smaller than the height
 of the first gravitational state $H_1$ (see Table \ref{Table1}) the
 flux of neutrons passing through the slit was indistinguishable from
 the background. As soon as the absorber position was set above $H_1$
 a rapid increase in the flux of neutrons was observed. An analogous
 increase, though less resolved, was observed for the slit heights
 close to the characteristic state height $H_2$. This "step-like"
 dependence faded almost completely for higher positions of the absorber, where the flux
increased practically monotonously.

 We will show here that such behavior of the neutron flux detected
 at the exit of the wave-guide is what one would expect from the
 qualitative treatment of neutron quantum motion in the gravitational
 field. In fact, the transversal motion of neutrons in the wave-guide
 can be described as a superposition of the neutron wave-guide transversal
modes:
 \[\Phi(z,t)=\sum_n C_n \psi_n(z)\exp(-iE_n t-\Gamma_n t/2) \]
 Here $\psi_n(z)$ represents the transversal states wave-functions, $E_n$
 the transversal self-energies and $\Gamma_n$ the widths of
 these states due to the neutron interaction with an absorber. The
 neutron flux, detected at the exit of the wave-guide is:
 \[
 F=\int_0^{\infty} |\Phi(z,\tau^{pass})|^2 dz
 \]

 The WKB approach can be proposed for the estimation of the
 widths of transversal states:
 \begin{equation}\label{Gsem}
 \Gamma_n=P_n\omega_n
 \end{equation}
  where $P_n$ is the probability of
 absorption of a neutron with energy $E_n$ by an absorber during a
 "one-time collision", while $\omega_n$ is the frequency of these
 collisions. The classical expression connecting the frequency of bouncing particle and
 its classical turning point $H_n$ is:
 \begin{equation}\label{FrCl}
 \omega_n=\frac{1}{2}\varepsilon_0\sqrt{\frac{l_0}{H_n}}
 \end{equation}

 We will use the following simple model for $P_n$. Namely, we will
 consider $P_n=1$ when the absorber height $H$ is below or equal $H_n$, so that
 a neutron can "touch" an absorber while it bounces above the mirror in
 the $n$-th state. If $H>H_n$ the probability is equal to
 the probability of tunneling through the gravitational barrier
 $P=D(E_n,H)$ \cite{GKK}. Such a probability has the following form
 in cases where $H\gg H_n$:

 \begin{equation}\label{Dgrav}
 P_n=D(E_n,H)\sim \exp \left[-\frac{4}{3}( (H-H_n)/l_0)^{3/2}\right]
 \end{equation}

 The spectrum of transversal states depends on the position of
 absorber $H$. As long as $H> H_n$ the first $n$ states can have a long
 enough lifetime to pass through the wave-guide:
 \begin{equation}\label{tlong}
 \tau_n^{long}=\frac{1}{ \omega_{n}}
 \exp\left(4/3(H/l_0-\lambda_n)\right)^{3/2}
 \end{equation}

  The lifetime of all other states with $E> E_n$ is approximately equal to the classical time of
flight of the
 particle with energy $E$ from the mirror to the absorber.
  We consider that such a lifetime is short compared to the passage time
 $\tau^{pass}$ through the wave-guide (which is ensured by the choice
 of length of the wave-guide $L$ and the horizontal flux velocity
 $V$) and their contribution to the detected flux is
 small as far as $\tau^{short}\ll \tau^{pass}$.

\begin{figure}
  \centering
 \includegraphics[width=115mm]{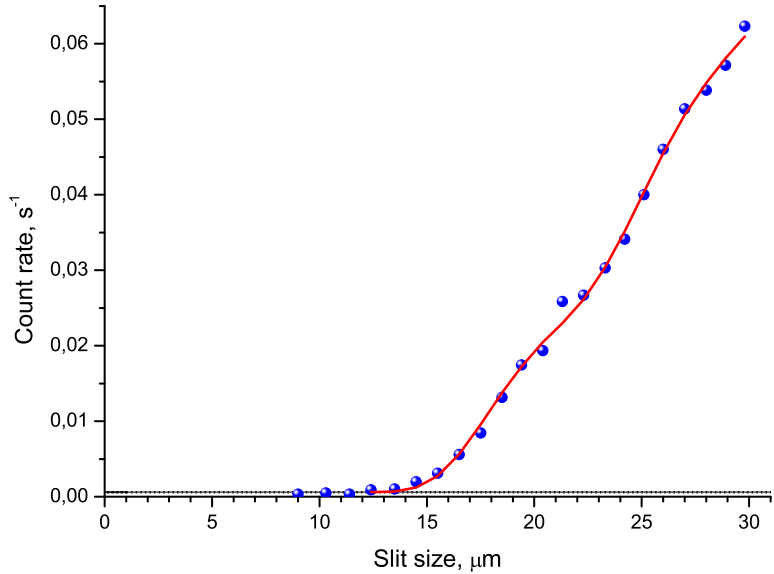}
  \caption{Neutron count rate at the exit of the wave-guide as a function of absorber position. Circles correspond to
   the experimental data,
  solid line correspond to the theoretical fit.}\label{Fig1}
 \end{figure}

  The measured neutron flux is:
 \begin{equation}\label{Fquant}
 F\simeq \sum_{n=1}^N|C_n|^2\exp(-\tau^{pass}/\tau_n^{long})
 \end{equation}
  Thus the measured flux exhibits a fast increase when absorber
  position $H$ is set close to $H_n$ due to the exponential increase of
  the $n$-th state lifetime (\ref{tlong}), which enables the passage of
neutrons in such a state through the wave-guide. (A more accurate
expression which includes the interference effects between decaying
states will be obtained in a later section). The expression
presented above was used to fit the experimental data:

  \begin{equation}\label{QFlux}
  F(H)=\sum_{n=1}^N A_n\exp(-\tau^{pass}/\tau_n^{long}(H))
  \end{equation}
  where $\tau_n^{long}(H)$ are defined by expression (\ref{tlong})
  with $\lambda_n$ used as free parameters, while
  $A_n=|C_n|^2$ were used to fit the "initial populations" of
  transversal states.  The fitted
values of $\lambda_n$ (taking to account the final accuracy of the
height calibration) are  in agreement with the expectation, as given
in Table I. The values of $A_n$ turned out to be equal, except
$A_1\simeq 0.7A_{n}$ with $n\geq 2$. The reason for the approximate
equality of the
  "initial populations" will be discussed
  in the later section. In Fig.\ref{Fig1} we show the experimental data (2002 year run \cite{Netal3}) and the
  results of the fit.

  When $H\simeq H_N$ and $N\gg1$ a large number of states passes through
the wave-guide. This number can be found from the WKB expression for
eigenvalues (\ref{WKB}):
  \begin{equation}\label{Nwkb}
  N^{WKB}=\frac{2}{3\pi}\left(\frac{H}{l_0}\right)^{3/2}+1/4
  \end{equation}
  Thus for big $N\gg 1$ the detected flux as a function of
  $H$ turns to be:
  \begin{equation}\label{Fwkb}
  F(H)\sim (H/l_0)^{3/2}
  \end{equation}

  The above mentioned WKB expression describes well the flux behavior
already for $N> 5$.
  The deviation of the measured flux from the above expression for small
$N$ is due to the quantum character
  (\ref{QFlux}) of the neutron motion in the gravitational field of Earth.
 Such a deviation (see Fig.\ref{Fig1}) is clearly seen for the first state when quantum
formula exhibit distinct threshold behavior at $H=H_1$.
 However the experimental possibility of resolving higher quantum
 states is restricted by the penetrability of the gravitational
 barrier. In fact the best resolution of the gravitational quantum
 states is achieved when the flux (\ref{Fquant}) has a step-like
 dependence on $H$. This means that the transition factor for given
 state $\exp(-\tau^{pass}/\tau^{long}_n(H))$ changes from the small
 value to unity in the range of absorber positions
 $H=\widetilde{H}_n\pm \delta_n$. The rate of such an increase is
 limited by the penetration probability through the gravitational
 barrier $D(E,H)$ (\ref{Dgrav}).

 For the clear resolution of different quantum states one needs
 $\delta_n\ll \widetilde{H}_{n+1}-\widetilde{H}_{n}$. Under the conditions of our experiment
 $\delta\simeq l_0$ and $\widetilde{H}_{n}\simeq H_n$. Such an
 estimation shows that for the highly excited gravitational states (with
practically $N\geq5$) the difference $H_{n+1}-H_n$ becomes
 comparable with the uncertainty $\delta$ and thus the step-like
 behavior of the flux is suppressed. To go
beyond the
 above-mentioned qualitative predictions of the resolution of
 gravitational states one needs to take into account details of
 the interaction of the neutron and the absorber. We will return to
 the discussion of the problem of the resolution of excited gravitational
 states in the later section.
 \section{Flat absorber}
 The simplest approach in which the properties of the
 absorber could be taken into account is a model of a flat absorber, characterized by the
 \emph{complex} Fermi potential. The simplification of the
 theory in the case of \emph{flat} absorber is due to the fact that, in
 such cases, motion in a transversal direction is independent of
 motion in a longitudinal direction within the wave-guide.

 \subsection{Passage of the neutron through an absorbing wave-guide}

 The Schr\"{o}dinger equation, which governs the wave-function
 $\Phi(x,z)$ of the neutron with total energy $E$ passing through the
 wave-guide is:
 \begin{equation}\label{SchWG}
 \left[-\frac{1}{2m}\frac{\partial^2}{\partial
 x^2}-\frac{1}{2m}\frac{\partial^2}{\partial
 z^2}+mgz+V(H,z)-E\right]\Phi(x,z)=0
 \end{equation}

 Here $x$ is the longitudinal variable, $z$ is the transversal
 variable, $V(H,z)=V_1(H,z)-iV_2(H,z)$ is the \emph{complex} Fermi
 potential of the absorber dependent on the absorber position $H$.

 It is convenient to introduce the transversal states $\psi_n(z)$,
 which are the eigenstates of the transversal Hamiltonian:

 \begin{equation}\label{trans}
 \left[-\frac{1}{2m}\frac{\partial^2}{\partial
 z^2}+mgz+V(H,z)-(\varepsilon_n(H)-i\Gamma_n(H)/2)\right]\psi_n(z)=0
 \end{equation}

 where $\varepsilon_n(H)-i\Gamma_n(H)/2$ are the \emph{complex}
 energy eigenvalues, dependent on absorber position $H$. It is worth
 noting that due to the presence of absorption (i.e. the imaginary
 component of $V(z,H)$)) the above mentioned transversal Hamiltonian
 is no longer self-adjoint. As a consequence, the eigenfunctions
 $\psi_n(z)$ are substantially complex and obey the bi-orthogonality
 condition (see \cite{BO} and references therein) :
 \begin{equation} \label{BiOrt}
 \int_{0}^{\infty}\psi_k(z)\psi_n(z)dz=\delta_{kn}
 \end{equation}
 As one can see the above expression differs from the standard
 orthogonality condition in the absence of complex conjugation.

  From the qualitative treatment of the previous section one can
 expect that the lifetime of the neutron in a transversal state
 $\psi_n(z) $ strongly depends on the absorber position $H$. The
 first $n$ lowest states such that $H_n\ll H$ are weakly affected by
 the absorber and practically coincide with gravitational states
 (\ref{Ai}). Their lifetime is large compared to the passage time
 $\tau^{pass}$. The states with $H_n \gg H$ are strongly distorted by
 the absorber. We will show that the corresponding lifetimes are
 short in comparison with $\tau^{pass}$ and these states totally
 decay before reaching the detector. Consequently, only states with
 rather small transversal energy and thus small width have a chance of
 exiting the wave-guide. When the absorber position $H$ is reaching
  one of the characteristic classical turning points $H_n$
 the corresponding state lifetime (and the wave-guide
 transition factor) undergoes fast changes with $H$, which allows us to
 monitor this quantum state in the overall flux at the exit of the
 wave-guide. To calculate the transition factors for the given state we
 expand the two-dimensional wave-function $\Phi(x,z)$ in the  set
 of basis functions $\psi_n(z)$:
 \begin{equation}\label{exp}
 \Psi(x,z)=\sum_n\chi_n(x)\psi_n(z)
 \end{equation}

 The functions $\chi_n(x)$ play the role of longitudinal wave-functions
 of neutrons in transversal state $n$ and can be found by
 substitution of (\ref{exp}) into the Schr\"{o}dinger equation
 (\ref{SchWG}) with the use of (\ref{BiOrt}):
 \begin{equation} \label{PlwaveEQ}
 \left[-\frac{1}{2m}\frac{\partial^2}{\partial
 x^2}+\varepsilon_n(H)-i\Gamma_n(H)/2-E\right]\chi_n(x)=0
 \end{equation}

 The solutions of (\ref{PlwaveEQ}) corresponding to the quasi-free longitudinal motion of neutrons in the wave-guide are:
 \begin{equation} \label{plwave}
 \chi_n(x)\sim\exp(ip_n x)
 \end{equation}
 Here $p_n=\sqrt{2m(E-\varepsilon_n(H)+i\Gamma_n(H)/2)}$ is the
 \emph{complex} longitudinal momentum.

 As we have already mentioned, only states with small transversal
 energy can reach the detector. In our case the full energy is much
 greater than the transversal energies:
 \[|\varepsilon_n(H)+i\Gamma_n(H)/2)|\ll E\]
 Thus we can write for momentum $p_n$:
 \[p_n\simeq\sqrt{2mE}(1-\frac{\varepsilon_n(H)-i\Gamma_n(H)/2}{2E})=P-\frac
{\varepsilon_n(H)-i\Gamma_n(H)/2}{V}\]
 where $P=\sqrt{2mE}$ and $V=P/m$. Due to the positive imaginary
 part of $p_n$, each of the longitudinal wave-function decays exponentially  with $x$
 inside the wave-guide:
 \[
 \chi_n(x)\sim\exp(i
 \frac{-\varepsilon_n(H)x+i\Gamma_n(H)x/2}{V})\exp(iPx)
 \]
  Taking into account that $\tau^{pass}=L/V$, we obtain for the
wave-function $\Psi(x=L,z)$
  at the exit of the wave-guide:
 \begin{equation}\label{exit}
 \Psi(x=L,z)=\exp(iPL)\sum_n C_n
 \exp(-i\varepsilon_n(H)\tau^{pass})\exp(-\frac{\Gamma_n(H)\tau^{pass}}{2})
\psi_n(z)
 \end{equation}

 Here $C_n$ are expansion factors, determined by the particular form of
 the wave-function at the entrance of the wave-guide:
 \begin{equation}\label{cn}
 C_n=\int_0^{\infty}\Psi(x=0,z)\psi_n(z)dz
 \end{equation}

 \subsection {Expansion factors}

  It is worth mentioning that strictly speaking $|C_n|^2$ cannot be
interpreted as the initial population of the certain state
$\psi_n(z)$. In fact, as long as the standard orthogonality
condition is not valid for the eigenfunctions of the
not-self-adjoint Hamiltonian $\langle \psi_n|\psi_k \rangle\neq
\delta_{nk}$ we
 have:
 \[\sum_n |C_n|^2 \neq \int_0^{\infty}|\Psi(x=0,z)|^2dz \]
 Consequently, to find the measured neutron flux at the exit of the
 wave-guide one has to take into account that:
\begin{equation}
\label{Fexit}
F=\int_0^{\infty}|\Psi(x=L,z)|^2=\sum_{n,k}C_n^{*}C_k\langle\psi_n
|\psi_k \rangle
\exp(-i(\varepsilon_k-\varepsilon_n)\tau^{pass})\exp(-\frac{(\Gamma_n+\Gamma_k)\tau^{pass}}{2})\neq
\sum_n |C_n|^2 \exp(-\Gamma_n\tau^{pass})
\end{equation}

 The appearance of the interference terms $C_n^*C_k\langle\psi_n
 |\psi_k \rangle$ is not surprising. In fact, the states $|\psi_k
 \rangle$ are \emph{not stationary} states with certain energy.
 Due to final decay width these states are in fact time-dependent and
 can be expressed as superpositions of \emph{stationary} states with
 \emph{certain} energy. The contribution of the mentioned
 interference terms to the flux can be interpreted as oscillating in
 time transitions with frequency
 $\omega_{nk}=\varepsilon_n-\varepsilon_k$ between the true
 stationary states.

 However for the observation of the interference terms above, a rather
narrow distribution of neutrons is required in the
 \emph{longitudinal} velocity. Should such longitudinal velocity
 distribution be broad, the interference terms are canceled after
 averaging over such a distribution and the "standard" expression
 for the flux is restored:
 \begin{equation}\label{Fstand}
  F=\sum_n |C_n|^2 \exp(-\Gamma_n\tau^{pass})
 \end{equation}

 Indeed, the interference terms $C_n^{*}C_k$ appear in the expression
 for the measured flux (\ref{Fexit}) multiplied by $\exp(i\omega_{nk}
 \tau^{pass})$. If the initial flux has distribution $f(V)$ in
 the longitudinal velocity, the contribution of the
 interference terms averaged over such a distribution would be:
 \[
 \int C_n^{*}C_k\langle\psi_n |\psi_k \rangle
 \exp(i\omega_{nk}\tau^{pass})\exp(-\frac{(\Gamma_n+\Gamma_k)\tau^{pass}}{2}
)f(V)
 dV
 \]
 In the case broad velocity distributions, such that:
 \[\frac{\Delta V}{V}\geq (\tau^{pass}\omega_{nk})^{-1} \]

 the contribution of the interference terms is canceled due to the
 fast oscillating term $\exp(i\omega_{nk}\tau^{pass})$. To observe
 the interference contribution between the first and second states, the
 velocity resolution in the conditions of our experiment should be
 better than 10\%. This limitation is less severe for excited states.

 Let us now turn to the problem of the initial "population" of the
 gravitational states where the initial flux has broad distribution
 in \emph{transversal} momentum. In such a case \cite{NP1} the modulus square of
 expansion coefficient $|C_n|^2$ can be found from the following
 equation:
\begin{equation}\label{Pop}
 |C_n|^2=\int\langle\psi_n|k \rangle \langle k|\psi_n \rangle  exp(-k^2/k_0^2) dk
 \end{equation}
 where $k_0$ is a characteristic width of the \emph{transversal} momentum
 distribution and we have used "bra-ket" notation for the matrix element
 $\langle\psi_n|k\rangle =\int\psi_n(x)\exp(ikx) dx$.

 It has been shown in \cite{AW,NP1,NP2} that if $k_0l_0\gg 1$ the
 squares of the amplitudes of the lowest states are practically equal:
 \[|C_n|^2\sim 1-o(\frac{1}{k_0l_0})\]
  In the conditions of our experiment the corresponding value $k_0l_0\simeq
 50$ and thus the approximation of a unified population of lowest states is
well justified.

 Indeed, having in mind fast
 oscillations of the integrand in (\ref{Pop}) $|C_n|^2$  becomes very small if $k >k_c$, where $k_c\simeq 1/
 H_n$ is the characteristic momentum of the gravitational state with
 spatial extension $H_n$.
  As far as the distribution over $k$ in the initial flux is
practically uniform for $k<1/H_n\ll k_0$
  the expression (\ref{Pop}) can be rewritten as:
 \[
 |C_n|^2=\int\langle \psi_n|k \rangle \langle k|\psi_n\rangle
  dk=\langle \psi_n|\psi_n \rangle =1
  \]
 and we return to the statement of the uniform distribution. It is
 worth mentioning that the same averaging over the initial transversal
 momentum distribution applied to an evaluation of the interference term
 $C_n^*C_k$ gives:
 \[C_n^*C_k=\langle \psi_n|\psi_k \rangle \]

 As we have already mentioned, this matrix element is nonzero for
 those states which are affected by the absorber and depends on absorber
 position $H$.
 Given the above arguments we can rewrite the expression for the measured
flux as a function of $H$ (\ref{Fexit}) after averaging over the
transversal momentum of the initial flux as:
 \begin{equation} \label{Fav}
 F(H)=\sum_{n=1}^{N}\exp(-\Gamma_n(H) \tau^{pass})+
 \sum_{n,k >n}^{N}2\mathop{\rm Re}\left(\langle\psi_n |\psi_k
 \rangle^2
 \exp(i\omega_{nk}(H)\tau^{pass})\right)\exp(-\frac{(\Gamma_n(H)+\Gamma_k(H)
)\tau^{pass}}{2})
 \end{equation}

 In the case of a broad longitudinal velocity distribution in the incoming
 flux, only the first term in this expression is important.

 %
 %
 %
 \subsection{Transition factor}
 Once the expression (\ref{Fav}) has been obtained, the problem of calculating
 the neutron flux at the detector position is transformed into the
 problem of calculating the eigen-energies $\varepsilon_n$ and
 their widths $\Gamma_n(H)$ of transversal states as a function of
 absorber position $H$.

 The realistic Fermi potential $V(H,z)$ of the absorber material is
 characterized by the depth of order $10^{-8} eV$ i.e much greater
 than the characteristic energy $10^{-12} eV$ of the lowest
 gravitational states. The diffusion radius $\rho$ of such a
 potential, i.e. the distance where the strength of potential rises
 from zero value in the free space to its final value
  inside the media, is much less than the characteristic gravitational
 wave-length $l_0$. In such a case the properties of the absorber can
 be precisely described by one parameter, namely the complex
 scattering length $a$, whose imaginary part accounts for the loss of
 neutrons due to absorption. (We use hereafter the following definition of the scattering length $a=\lim_{k\rightarrow 0}(1-S)/(2ik)$, where
 $k$ is neutron momentum and $S$ is the reflected wave amplitude). An analytical equation for the
 eigen-energies of neutrons bouncing in the gravitational field
 between mirror and absorber which is positioned at distance $H$ above the
 mirror can be derived. We refer the reader to
 Appendix A for the details and present here the final expression for the
 eigenvalues $\lambda_n(H)$:
 \begin{equation} \label{EnEQAbs}
 \frac{\mathop{\rm Ai}(-\lambda_n)}{\mathop{\rm
  Bi}(-\lambda_n)}=\frac{\mathop{\rm
Ai}(H/l_0-\lambda_n)-\tilde{a}/l_0\mathop{\rm
Ai'}(H/l_0-\lambda_n)}{\mathop{\rm
Bi}(H/l_0-\lambda_n)-\tilde{a}/l_0\mathop{\rm
  Bi'}(H/l_0-\lambda_n)}
  \end{equation}
 Here $\tilde{a}=a-H$ plays the role of "the scattering length on the
 diffuse tail" of the potential. Let us note that this expression is
 valid for \emph{any } Fermi potential with a small diffuse radius,
 as long as the corresponding scattering length $\tilde{a}$ is small
 compared to the gravitational wave-length $l_0$. If the
 position of absorber $H\gg H_n$ the right-hand side of the
 equation (\ref{EnEQAbs}) becomes exponentially small:
 \[\frac{\mathop{\rm Ai}(-\lambda_n)}{\mathop{\rm
 Bi}(-\lambda_n)}\simeq
  \frac{1}{2}\exp\left[-\frac{4}{3}( (H-H_n)/l_0)^{3/2}\right]\left(1+2\sqrt{(H-H_n)/l_0}\frac{\tilde{a}}{l_0}\right)
\]

 One can easily recognize in the right-hand side exponent the
 penetration probability through the gravitational barrier. Taking
 into account the smallness of such a probability one can find the
 correction to the gravitational eigenvalue $\lambda_n$ due to the
 small, but nonzero, possibility of penetration under the gravitational
 barrier to the absorber.
 \begin{equation}\label{Corrlamb}
 \Delta \lambda_n=-\frac{\mathop{\rm Bi}(-\lambda_n^0)}{2\mathop{\rm
  Ai'}(-\lambda_n^0)}\exp \left[-\frac{4}{3}(
(H-H_n)/l_0)^{3/2}\right]\left(2\sqrt{(H-H_n)/l_0}\frac{\tilde{a}}{l_0}+1\right)
 \end{equation}
 where $\lambda_n^0$ are unperturbed eigenvalues determined by
 (\ref{lambda0}). The width of the $n$-th state due to penetration
 under the gravitational barrier and absorption turns out to be:
 \begin{equation}\label{Gfl}
 \Gamma_n\simeq 2\frac{|\mathop{\rm
 Im}a|}{l_0}\varepsilon_0\sqrt{\frac{l_0}{H_n}}\sqrt{(H-H_n)/l_0}\exp
 \left[-\frac{4}{3}( (H-H_n)/l_0)^{3/2}\right]
 \end{equation}
 We have used in the derivation of this expression the semiclassical
 approximation for the Airy function. The physical sense of this
 expression for the decay rate becomes clear after comparison
 with the semiclassical expressions (\ref{Gsem}). Taking into account
 the expression (\ref{FrCl}) for the classical frequency $\omega_{n}$
 we can rewrite (\ref{Gfl}) as:
 \[\Gamma_n\simeq 4\frac{|\mathop{\rm
 Im}a|}{l_0}\omega_{n}\sqrt{(H-H_n)/l_0}\exp \left[-\frac{4}{3}(
 (H-H_n)/l_0)^{3/2}\right] \]

  The neutrons  penetrate through the gravitational
barrier into
 the absorber, the corresponding probability (\ref{Dgrav}) is exponentially small.
  This probability is multiplied by the classical
 bouncing frequency in given state $n$. The properties of the
 absorber itself appear in the above expression through the ratio
 $4\frac{|\mathop{\rm Im}a|}{l_0}\sqrt{(H-H_n)/l_0}$. Later we will
 show that it coincides with a general expression for the absorption
 probability of slow quantum particles on the short-range absorbing
 potential. Thus the intuitive formula (\ref{Gsem}) we used before
 is justified.

 One can introduce the characteristic absorption time:
 \[\tau^{abs}_n=\frac{l_0}{2|\mathop{\rm
 Im}a|\varepsilon_0}\sqrt{\frac{H_n}{l_0}}\]



 For efficient absorption one needs
 $\tau^{pass}/\tau^{abs}_n\gg1$, which puts the following requirement for the
 imaginary part of the scattering length $\mathop{\rm Im}a$:
 \[|\mathop{\rm Im}a|\gg
 \frac{l_0}{2\varepsilon_0\tau^{pass}}\sqrt{\frac{H_n}{l_0}}\]

 In the conditions of our experiment ($\tau^{pass}=2\times 10^{-2}$ s) the
 above requirement means that
 \begin{equation} \label{amin}
 \mathop{\rm Im}a\gg 0.05 l_0\sim 0.3 \mu m
 \end{equation}

 (Let us mention that at the same time the scattering length
 approximation used above is valid only for values of $|\tilde{a}|\ll
 l_0$)

 This treatment shows that for the clear resolution of
 quantum states the most favorable absorbers are those with the largest
 possible scattering length.

 Let us see how the scattering length discussed above is connected to
 the properties of the absorber's  Fermi potential,
namely its complex depth  and the diffusion radius $\rho$. We will
study the case of the \emph{complex} potential of the Woods-Saxon
type:
 \begin{equation}
 \label{Fermipot} V(z,H)= \frac{U\exp(-i\varphi)}{1+\exp((H-z)/\rho)}
 \end{equation}

In the following $U>0$.
 It can be shown \cite{Flugge} that the scattering length on such a
 potential is given by:
 \begin{equation} \label{WSsl}
 a=H-\frac{1}{\kappa}+2\rho\left(\gamma+
 \frac{\Gamma'(1+\rho\kappa)}{\Gamma(1+\rho\kappa)}\right)
 \end{equation}
 Here $\gamma\approx0.577$ is the Euler constant,
 $\kappa=\exp(-i\varphi/2)\sqrt{2mU}$ and $\Gamma(x)$ is the
 Gamma-function.

  An important limiting case is the case of the deep complex
 Fermi-potential, namely $\rho|\kappa|\gg 1$ and $\rho|\mathop{\rm
 Im}\kappa|\gg 1$. It follows from (\ref{WSsl}) that in such a case:
 \begin{equation}\label{Exp}
 \mathop{\rm Im}a=-\rho \varphi
 \end{equation}
  and is \emph{independent} of the depth $U$ of the complex Fermi potential.
 It has been shown in \cite{KPV} that such behavior of the imaginary
 part of the scattering length is universal for deep complex
 potentials with the exponential tail. For such a strong absorbing
 Fermi potential the neutron is completely absorbed on the tail of
 the complex Fermi potential; the properties of the inner part
 of the absorber therefore lose their importance. The only way to increase
 the scattering length in such a limit is to increase the diffuseness
 $\rho$.

 In an other limiting case, when the diffuseness is so small that
 $\rho\kappa\ll 1$ the
  scattering length becomes:
 \[\mathop{\rm Im}a=-\mathop{\rm
Im}\frac{1}{\kappa}-\frac{\pi^2}{3}\rho^2\kappa \]
 The leading term in the above expression is $-1/\kappa$. One can
 check that it coincides with the scattering length, characterizing
 the low-energy reflection from the step-like potential
 $U\exp(-i\varphi)\Theta(z)$. Indeed, in the case of $\rho\kappa\ll 1$ the
 neutron can penetrate through the narrow exponential tail of the
 complex potential into its core without significant losses. One can see from the above expression
that for
 \emph{weak} absorbers with depth $U\sim \varepsilon_0$ the
 imaginary part of the scattering length becomes as large as $l_0$.

 It is important to mention here that the absorption of the
 ultra-cold neutrons by the complex potential is closely related to
 the so-called quantum reflection \cite{QR,QR1} of ultra-slow neutrons
 from the fast changing complex Fermi potential. The reflection
 probability $R$ in case of slow quantum particles \cite{LL} impinging
 on the absorber at normal incidence with momentum $k$ can be written as
 follows:
 \begin{equation}\label{QR}
 R=1-4k|\mathop{\rm Im} a|
 \end{equation}
 while the absorption probability $P$ is:
 \begin{equation}\label{QA}
 P=1-R=4k|\mathop{\rm Im} a|
 \end{equation}

  The smaller the $\rho$ (and $|\mathop{\rm Im} a|$) the better the
reflection and the weaker the absorption.
 One can see that the limit of $\kappa\rho \rightarrow 0$,
 $p/\kappa\rightarrow 0$ corresponds to the case when the absorber is
 replaced by an absolutely reflecting mirror. The above mentioned high
 reflectivity of a fast changing potential is a general quantum
 mechanical property.

 The numerical calculations verify the above conclusions.

 The values $U$, $\rho$ and $\phi$ of potential (\ref{Fermipot})
 were chosen to be $U=10^{-8}$ eV,  $\rho=1$ $\mu m$ and $\phi=3\pi/4$ (this corresponds to attractive potential with absorption).

 \begin{figure}
  \centering
 \includegraphics[width=75mm]{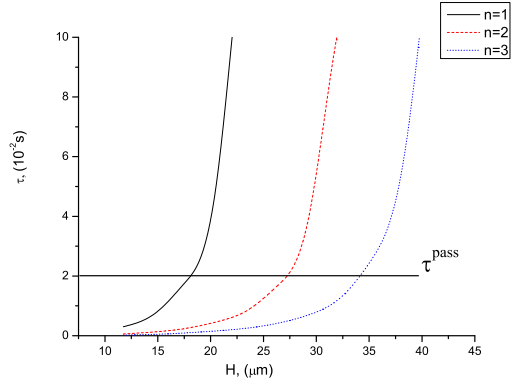}
  \caption{Lifetime of the first three gravitational states as a function
of absorber position in a typical arrangement of our
experiment}\label{Fig2}
 \end{figure}
 \begin{figure}
  \centering
 \includegraphics[width=75mm]{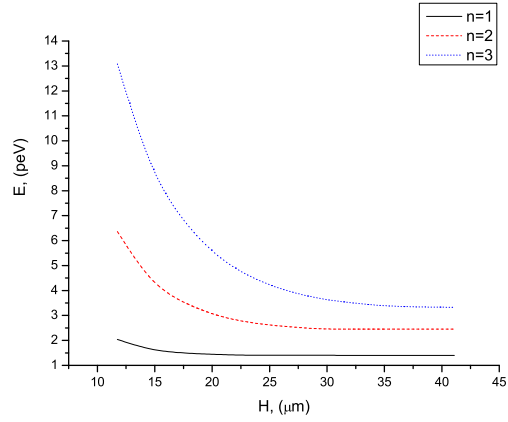}
  \caption{Energy of the first three gravitational states as a function of
absorber position in an arrangement typical for our
experiment}\label{Fig3}
 \end{figure}
 \begin{figure}
  \centering
 \includegraphics[width=75 mm]{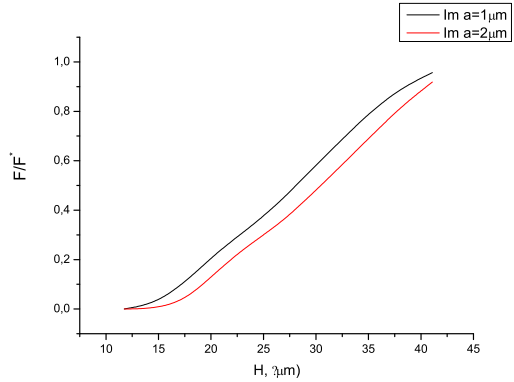}
  \caption{The relative neutron flux as a function of absorber position for
different absorber difuseness in an arrangement typical for our
experiment.
  $F^*$ is the flux calculated at absorber position $H=45$ $\mu m$ and
diffuseness $\rho=1$ $\mu m$.}\label{Fig4}
 \end{figure}
 In Fig.\ref{Fig2} we plot the lifetimes $\tau_n$ of the first 3
 states as a function of absorber position $H$. In Fig.\ref{Fig3} we
 show the corresponding evolution of the real part of energy
 $\varepsilon_n(H)$. One can see the fast increase in the lifetimes
 at certain values of $H$, close to $H_n$, in agreement with the
 qualitative predictions of the previous section. The real part of
 the energy quickly approaches its limiting value equal to the energy
 of the gravitational state when $H> H_n$.

 As long as these states are treated separately the clear evidence of
 the fast changes in the lifetime, as a function of absorber position
 can be seen. However, the overall plot of flux intensity
 Fig.\ref{Fig4}, where all these states are taken into account
 simultaneously shows that the step-like dependence is suppressed, except for the first step at $H=H_1$
and partially for
 the second step.

 To achieve much higher absorber efficiency the diffuse radius
 $\rho$ should be significantly increased. Another way is to reduce
 the depth of absorber Fermi potential to the level  of  $10^{-12}$
 eV, the characteristic scale of gravitational states energies.


 Absorbers with optimal parameters can be obtained if their surface is
corrugated. In fact such an absorber was used in the experimental
set-up. The zone of such a corrugation can be
 considered as a low density media with an extended diffuse radius of Fermi
 potential. In the following we will study the neutron passage
 through an absorber with a rough surface. We will show, however, that
 the main loss mechanism in such a case is due to non-specular
 reflections from the rough edges of absorber.

 \subsection{Zero gravity experiment}

  We will study here the important case of the
 neutron passage through the wave-guide formed by the mirror and the
 absorber in the \emph{absence} of the gravitational field. The
 case is interesting from two points of view. On the one hand, a comparison
of the transition factors with and without gravity clearly shows the
role
 of the latter \cite{Sweden,Answer}. On the other hand the "zero"
 gravity experiment (which simply means the installation of the
 mirror and the absorber \emph{parallel} to the gravitational field)
 enables independent measurement of the mirror and absorber
 properties.

 Let us first mention that the neutrons' motion transversal to the
direction of
 the mirror (and the absorber) is quantized. Neutron states of this type,
localized between the mirror and the absorber, will be
 referred to as "box-like". However due to the loss of neutrons
 inside the absorber such states are no longer stationary states;
 they are quasi-bound states with finite life-times (width). The
 existence of quasi-bound states in the presence of an absorber is a
 consequence of the phenomenon mentioned above as quantum reflection
 from the fast changing absorber Fermi-potential. In fact the
 partial reflection of neutron waves from the absorbing potential leads to
the formation of the standing wave (i.e. quasi-bound state). The
more efficient
 the absorber the smaller the amplitude of the reflected wave
 and the shorter the neutron life-time. In the case of full absorption
 of the neutron wave (which means that the amplitude of the reflected
 wave is exactly zero) no quasi-bound state can exist.

 With these remarks we can now turn to the calculation of the neutron
 flux through the wave-guide:
 \begin{equation}\label{Fstand1}
  F=\sum_n |C_n|^2 \exp(-\Gamma_n\tau^{pass})
 \end{equation}

 Here $n$ is the quantum number of the  quasi-bound box-like state. In the above
 expression we neglect for the moment the contribution of the
 interference terms. As we have shown, this is possible when the
 \emph{longitudinal} velocity distribution is rather wide. In the
 following we will also assume a wide distribution of the incident flux
 over transversal momentum (orthogonal to the mirror and absorber).
 We have already established that in such a case the first $N_h$
 states are populated homogenously. The number $N_h$ of homogenously
 populated states can be estimated from the condition that
 the characteristic momentum of the box-like state $k_c \sim n/H$ is
 equal to the spread of the transversal momentum distribution $k_0$
 in the incident flux:
 \begin{equation}
 N_h \simeq k_0H \label{Nhom}
 \end{equation}

 Let us now turn to the calculation of the widths of certain neutron
 states, confined between mirror and absorber. As in the case of the
gravitation states we assume that the absorber Fermi
 potential can be characterized by a complex scattering length $a$,
 which is possible when $k_n\rho\ll 1$, (where $k_n=\sqrt{2mE_n}$ is
 the neutron momentum in given box-like state with energy $E_n$). To
 obtain the complex energies of the box-like states we note that the
 neutron wave function in the region where the absorber Fermi
 potential can be neglected is:
 \[\Psi_b(z) \sim \sin(k_n z) \]

  Such a wave function can be matched with the asymptotic form of the
neutron wave-function inside the absorber at distances
  $H-1/k_n\ll z\ll H-\rho$, where absorber potential vanishes. The general
asymptotic form of the wave-function in this region is:
 \[\Psi_a(z) \sim 1+\frac{H-z}{\tilde{a}}\]
 where $\tilde{a}=a-H$ is the "diffuse tail" scattering length. The
 matching of the wave-function and its derivative leads to:
  \begin{eqnarray}\label{Kn}
  k_n&=&\frac{\pi n}{H-\tilde{a}} \\\label{Ebox}
  E_n&\approx &\frac{\pi^2n^2}{2mH^2}+\frac{2\pi^2n^2\mathop{\rm Re}
\tilde{a}}{2mH^3} \\ \label{Gbox}
 \Gamma_n&\approx &4E_n\frac{|\mathop{\rm Im} a|}{H}=
 4\frac{\pi^2n^2|\mathop{\rm Im} a|}{2mH^3}
 \end{eqnarray}

  We will show that the dependence (\ref{Gbox}) will play a crucial
  role in establishing the wave-guide transition factor dependence on
  $H$. Let us note here that the expression for the
  width of the box-like state (\ref{Gbox}) is a consequence of the quantum
reflection
  from the fast
  changing tail of the absorber Fermi potential. To see how the quantum
reflection phenomenon is connected
  to the width of the box-like state let us return to
  the semiclassical expression for the loss rate:
  \[\Gamma\sim \omega_n P\]
  where $\omega_n$ is the classic frequency of collisions with the
  absorbing wall and $P$ is the probability of absorption in a "one
  touch" collision. The expression for the collision frequency with
  \emph{one of two} walls is
  \[\omega_n=\frac{v_n}{2H}=\frac{k_n}{2mH}=\frac{\pi n}{2mH^2}\]
  The probability of absorption $P$ (\ref{QA}) turns to be:
  \begin{equation}\label{QA1}
  P=1-R=4k_n|\mathop{\rm Im} a|=\frac{4\pi n|\mathop{\rm Im} a|}{H}
  \end{equation}
  Combining the above results for the frequency $\omega_n$ and absorption
  probability $P$
  we return to the expression (\ref{Gbox}). The quantum properties of
neutron motion appear here through the energy dependence of the
absorption probability (\ref{QA1}) and quantization of the box-state
energy (momentum).

  Integrating the results for $C_n$ and $\Gamma_n$ into the expression
  for the flux (\ref{Fstand}) we obtain:
  \begin{equation} \label{Tbox}
  F= F_0\sum_n \exp\left(-4\frac{\pi^2n^2|\mathop{\rm Im}
  a|}{2m H^3}\tau^{pass}\right)
  \end{equation}
  Here $F_0$ is the normalization constant, characterizing the
  intensity of initial flux.

  One can see that the number of states passed through the wave-guide
  is obtained from the condition:
  \[\Gamma_n\tau^{pass} \simeq 1\]
  which gives
  \[N^{pass}\simeq \frac{H^{3/2}\sqrt{2m}}{2\pi\sqrt{|\mathop{\rm Im}
  a|\tau^{pass}} }\]

  Hereafter we expect that the number of homogeneously populated states
$N_h$ (\ref{Nhom}) is greater than $N^{pass}$ :
  \[k_0H\geq N^{pass}\]

  From the expression (\ref{Tbox}) it follows that the wave-guide
transition coefficient is determined by the characteristic
\emph{absorption constant}:
  \[\xi=4\frac{\pi^2\tau^{pass}|\mathop{\rm Im} a|}{2mH^3}\]
  which is connected with the number of states passed through the
  wave-guide via:
  \[N^{pass}\simeq1/\sqrt{\xi}\]

  There are two important limiting cases.

 The first case, which we call the \emph{"strong absorption"} limit
$\xi >1$, means that a maximum of one state only can pass through
the wave-guide, i.e.:
  \[ N^{pass}\simeq1/\sqrt{\xi} \leq 1 \]

  In this case the neutron flux  is a rapidly
increasing function of $H$:
 \begin{equation}\label{StrongAbs}
 F\approx F_0\exp\left(-4\frac{\pi^2|\mathop{\rm Im}
 a|}{2mH^3}\tau^{pass}\right)
 \end{equation}

  The opposite case, which we call the \emph{"weak absorption"}
  limit $\xi\ll 1$
  means that a large number of states can pass through the absorber:
 \[N^{pass}=1/\sqrt{\xi} \gg 1\]

  In this case the summation in expression (\ref{Tbox}) can be substituted by
  integration, which gives:
 \begin{equation}\label{WeakAbs}
 F\approx F_0\frac{H^{3/2}\sqrt{2m}}{2 \sqrt{2\pi|\mathop{\rm Im}
  a|\tau^{pass}}}
 \end{equation}

  It is worth mentioning that $H^{3/2}$-dependence is a consequence of the
quantum threshold behavior  that determines the energy
 dependence of the absorption probability of ultra-cold neutrons
 (\ref{QA}). This expression is valid in the so-called
 \emph{anti-classical limit} $k_n\tilde{a}\ll 1$. In the opposite
 case, when $k_n\tilde{a}> 1$ the absorption probability energy
 dependence differs from (\ref{QA1}). In particular, if the
 absorption occurs with unit probability for each collision
 $P\simeq 1$ it is easy to establish:
 \[ \Gamma_n\approx \pi
 n/(2H) \]
  The substitution of this expression into (\ref{Fstand}) results in the
$H^2$ dependence of the flux instead of $H^{3/2}$. Note that the
large value of the absorption probability of ultra-slow neutrons can
be achieved only if there is a large imaginary part of the
scattering length $k_n|\mathop{\rm Im}
 a| \gg 1$.

 Let us also note that we restrict ourselves with the condition of a
 homogeneous population of box-like states $k_0H\geq N^{pass}$,
 so far $\xi$ cannot be smaller than:
 \[\xi_{min}=\frac{1}{k_0^2H^2}\]

 Obviously, in the limit of very small
$\xi \ll \xi_{min}$, when absorption can be fully neglected, the
flux passed through the slit starts to be proportional to the slit
size:
 \[F\sim H\]
 which means that all the neutrons that enter the slit pass through
 it without losses. So far, depending on the efficiency of the absorber
 one can get different flux dependence on the slit size $H$.

 A comparison with the gravitational case results in the following
 conclusions.

 First, for small slit sizes, both flux curves, seen as a function
 of $H$ manifest fast increases in the vicinity of the characteristic
 value $H_c$. However in the case of gravitational states this critical
 slit size is determined by the "height" of the ground gravitational
 state $H_c\simeq H_1$, while in the case of zero gravity it is fully
 determined by the properties of the absorber, namely the imaginary part
 of the scattering length $|\mathop{\rm Im}
  a|$ and the passage time $\tau^{pass}$:
 \[H_c=(2\pi^2\tau^{pass}|\mathop{\rm Im} a|/m)^{1/3}\]

  Secondly, in the presence of gravitation the flux exhibits a step-like
dependence on $H$ with increasing slit size, and tends to $H^{3/2}$
 dependence for large $H$. Such step-like behavior is more
 pronounced for larger diffuseness of the absorber (larger
 $|\mathop{\rm Im} a|$) or for longer passage time. In case of zero
 gravity the flux increases with $H$ monotonously. Such an increase
 has power law dependence in the limit of large $H$. Depending on the
 absorber efficiency the corresponding exponent can vary from $2$
 (full absorption) to $1$ (no absorption).
 \begin{figure}
  \centering
 \includegraphics[width=75mm]{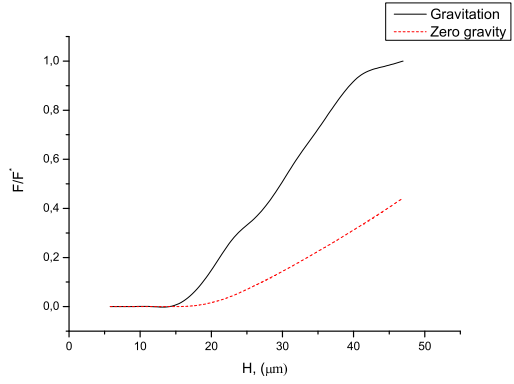}
  \caption{The relative neutron flux in the presence of, and in the absence
of, gravity. $F^*$ is the flux with gravity calculated for absorber
position $H=45$ $\mu
  m$,
  $|\mathop{\rm Im} a|= 2$ $\mu m$ and $\tau^{pass}=0.02$ s. }\label{Fig5}
 \end{figure}

 We plot on Fig\ref{Fig5} the neutron flux in the presence of and in the
 absence of gravity for $|\mathop{\rm Im} a|=2$ $\mu m$ and
 $\tau^{pass}=0.02$ s.

 \subsection{Inverse geometry experiment}
 Another way to clarify the gravitational effects and to measure
 the efficiency of the absorber is to exchange the position of
 the mirror and the absorber in the experimental setup. Here we will
 study such an inverse geometry experiment, in which the absorber
 is placed below and the mirror above.

 First we will study the modification of the gravitational energy
 values due to the interaction with an "absorbing mirror". As we have
 already shown, as long as the distance $\rho$ where absorption takes
 place is much smaller than the gravitational wave-length $l_0$, such
 an interaction can be characterized by only one parameter, namely
 the scattering length $a\ll l_0$, regardless of certain details of
 the absorber Fermi potential.

 The modification of the eigenvalues $\lambda_n$ due to the
 interaction with an "absorbing mirror" can be obtained by matching
 the wave-function of the neutron, reflected from the absorber,
 which
 large $z$ asymptotic form ($z\gg a$) in the case of small neutron
 energies can be written as follows:
 \[
 \psi(z)\sim 1-z/a
 \]
 with the gravitational wave-function $\mathop{\rm
 Ai}(z/l_0-\widetilde{\lambda})$, where $\widetilde{\lambda}$ is a
 modified eigenvalue. We take into account that in the matching region
 $z/l_0\ll 1$ we obtain the following equation for
 $\widetilde{\lambda}$:
 \begin{equation} \label{lambda}
 \frac{\mathop{\rm Ai}(-\widetilde{\lambda})}{\mathop{\rm
 Ai}'(-\widetilde{\lambda})}=-a/l_0
 \end{equation}

 As far as $|a/l_0|\ll1$ we get the following expression for the modified eigenvalues
$\widetilde{\lambda}_n$ accurate up to the first order of small
parameter $|a/l_0|$:
 \begin{equation}\label{lambnew}
 \widetilde{\lambda}_n=\lambda_n+a/l_0
 \end{equation}
  From the above equation we obtain the following modified energy levels:
 \begin{eqnarray}
 E_n&=&\varepsilon (\lambda_n+\mathop{\rm Re}a/l_0) \\
 \Gamma_n&=& 2\varepsilon \frac{|\mathop{\rm
 Im}a|}{l_0}=2mg|\mathop{\rm Im}a|\label{Ggrav}
 \end{eqnarray}

 If we use the expression (\ref{Exp}) for the scattering length on
 the deep ($2\rho \sqrt{2mU}\gg 1$) imaginary ($\varphi=\pi/2$)
 exponential potential, we obtain for the width of the gravitational
 state:
 \begin{equation}\label{Gexp}
 \Gamma^{inv}=mg\pi \rho
 \end{equation}

  One should mention that the width of the gravitational state
 (\ref{Ggrav}) is independent of the energy (for such states that
 $\sqrt{2mE_n}\rho \ll 1$). This can be easily explained by the
 following simple arguments. The frequency of the neutron bouncing
 above the surface in the gravitational field is $\omega \sim
 1/\sqrt{E}$, while the probability of the absorption
 $P=4k|\mathop{\rm Im} a| \sim \sqrt{E}$. Combining these two
 variables we get the energy-independent expression for the width
$\Gamma=\omega
  P$.
  This means that all the gravitational states which are not affected by
the upper mirror ($H_n\ll H$) decay at the same rate
  (\ref{Gexp}). The corresponding lifetime in case of $\rho=1$ $\mu
  m$ is

  \[ \tau\simeq 1.7 \times 10^{-3} \mbox{ s } \]

 which is much smaller than the passage time $\tau^{pass}=0.02$ s.
 The expression (\ref{Ggrav}) manifests the very important property
 of the neutron bouncing above the "absorbing mirror", namely the
 factorization of gravitational properties, which appears through
 factor $mg$ and the absorber properties, characterized by
 $|\mathop{\rm Im}a|$.

 Let us now understand the behavior of the transversal states with much
higher energy $E\gg mgH$. For such high energies the influence of
the gravitational field can be neglected (for neutron motion between
mirror and absorber). The corresponding states can be treated as the
previously studied
  "box-like" states of the free neutron, confined between the absorber and the
  mirror; their widths are given by (\ref{Gbox}).

 As long as we study the transversal states with $E_n\gg mgH$ their
 lifetimes are much smaller than those of the gravitational states already
considered, and their contribution to the neutron flux at the exit
of the wave-guide can be neglected.

 The two limiting cases studied above naturally follow from the equation
for the eigenvalues for the inverse geometry experiment (see
Appendix A for details of the derivation):
  \begin{equation}\label{InvEq}
  a\left[\mathop{\rm Ai}(H/l_0-\lambda_n)\mathop{\rm
  Bi'}(-\lambda_n)-\mathop{\rm Ai'}(-\lambda_n)\mathop{\rm
  Bi}(H/l_0-\lambda_n)\right]=\mathop{\rm Ai}(-\lambda_n)\mathop{\rm
  Bi}(H/l_0-\lambda_n)-\mathop{\rm Bi}(-\lambda_n)\mathop{\rm
  Ai}(H/l_0-\lambda_n)
  \end{equation}



\begin{figure}
  \centering
 \includegraphics[width=75mm]{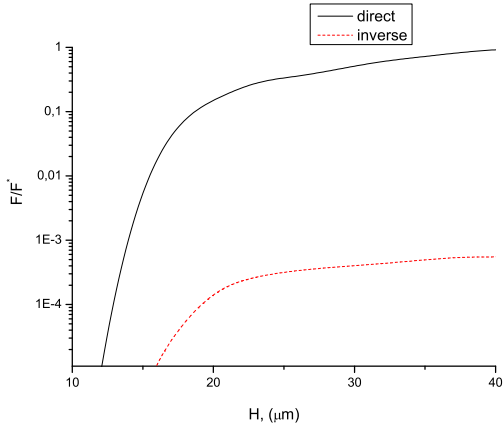}
  \caption{The relative neutron flux in the direct and inverse geometry
experiment. $F^*$ is the flux in the direct geometry case,
calculated for absorber position $H=45$ $\mu
  m$,
  $|\mathop{\rm Im} a|= 2$ $\mu m$ and $\tau^{pass}=0.02$ s.}\label{Fig6}
 \end{figure}

 We come to the conclusion, that, with mirror position $H\gg H_1$,
 the measured neutron flux is mainly determined by the gravitational
 states passed through the wave-guide such that $H_n<H$. The number
 of such states is given by (\ref{Nwkb}) and thus the dependence of
 the flux on $H$ is given by (\ref{Fwkb}). The ratio of the fluxes in the "direct" and in
 the inverse geometry experiment turns to be:
 \begin{equation}\label{FluxRatio}
 F_{inv}/F_{dir}\simeq\exp(-2mg|\mathop{\rm Im}a| \tau^{pass})
 \end{equation}
The results of comparison of neutron fluxes  in direct and inverse
geometry experiment is shown in
 Fig.\ref{Fig6}

  This difference in the fluxes clearly shows the role of gravitation
  in the passage of the neutrons through the wave-guide. On the other hand it
  enables us to measure the efficiency of the absorber.
  It is also interesting to note that if $|\mathop{\rm Im}a|$ is
  known by independent measurement (e.g. from the zero gravity
  experiment discussed above), the measurement of the lifetime of
  neutrons bouncing in the low gravitational states above an absorbing
  surface will give direct access to the \emph{gravitational} mass of neutron
  $m$ and will allow us to apply the quantum equivalence principle test.

  The inverse geometry measurements were performed
  during one of the first runs of neutron gravitational states
  experiment \cite{Netal2}.
  The obtained results verify strong suppression of the flux in inverse geometry experiment case
  in  agreement with
  (\ref{FluxRatio}).

 \section{ Rough surface absorbers}

 The previous analysis shows that in order to increase the efficiency
 of flat absorbers one needs to use substances either with a Fermi
potential of large
 diffuse radius or of very small depth ($U\sim 10^{-12}$ eV). The
construction of such absorbing materials is rather problematic. An
alternative way to increase absorber efficiency is to use an
absorber with a rough surface. In
 the wave-guide experiments an absorber with a rough surface was
 used with a roughness amplitude of about $2$ $\mu m$. In this section
 we will study the role of roughness in the neutron loss mechanism.

 \subsection{ Effective potential approach}

 The rigorous study of neutron interactions with a rough surface
 requires solving the two-dimensional problem, where the
 neutron-surface interaction is described by a rather complicated
 function $V(x,z)$. The radical simplification of such a problem is
 possible via introduction of \emph{effective} one-dimensional
 potential $V_{eff}(z)$ \cite{AW}. The simplest assumption enabling us to
 calculate such a potential is the following. We expect that the
 \emph{longitudinal} kinetic energy of neutrons $p_0^2/2M$ is
 sufficiently superior to the characteristic value of the Fermi
 potential $V(x,z)$ of rough edges. The first order Born
 correction to the longitudinal kinetic energy of neutrons due to the
 interaction with the rough edges would then be:

 \begin{equation}\label{Vef}
 \Delta E(z)=\frac{1}{L}\int V(x,z) dx
  \end{equation}
 where the "normalization length" $L$ is selected to be much greater than
 characteristic correlation length of roughness. This correction to the
longitudinal energy, being a function of $z$, plays the role of
effective potential $V_{eff}(z)=\Delta E(z)$ in the equation for the
neutron transversal motion:

 \[-\frac{1}{2m}\frac{d^2}{dz^2}\varphi (z)+\Delta
 E(z)\varphi(z)=(E-p_0^2/2m)\varphi(z) \]

 The physical meaning of expression (\ref{Vef}) is transparent; it is
 the potential of media with reduced density. In particular if one
 models the roughness by the periodic gratings with $z$ dependent
 width $d(z)$ and period $L$, than the effective Fermi potential is
 $V_{eff}(z)= U d(z)/L$, where $U$ is the corresponding Fermi
 potential of  flat surface. The benefit of this approach is the
 ability to connect the one-dimensional effective Fermi potential
 with  averaged shape properties of roughness and  realistic
Fermi
 potential of  absorber substance. We will not take this case any
 further, since the main results have already been discussed in the section
 devoted to the flat absorber.

 The above approximative model can be justified for the longitudinal
 energies of neutrons much higher than the Fermi potential of rough
 edges. A very important effect, which is not taken into account in
 this simplified approach is the possibility of non-specular
 reflections, i.e. the energy exchange between the horizontal and
 vertical motion of the neutrons. (They appear in the second order Born
 approximation). In the following we develop the non-perturbative
 formalism in which such effects would be taken into account.

 \subsection {A time-dependent model for the neutron loss mechanism}
 In the previous analysis we found that only those neutrons which
 have sufficiently small transversal energy do not penetrate through
 the gravitational barrier into the absorber and thus are not
 absorbed in the wave-guide. The role of the absorber's roughness is to
 transfer a significant portion of longitudinal energy into
 transversal energy during non-specular reflection from the rough
 edges. Thus the neutron interaction with the rough surface absorber results
 in mixing of states with different transversal energies. As long as
 the states with large transversal energy have very small lifetimes,
 such a mixing results in a loss of neutrons. Here we will study this loss
mechanism within the time-dependent model.

 We will study the neutron passage through the wave-guide in the
 frame, moving with horizontal velocity $V$ of the incoming flux (we
 suppose that this velocity is well defined). The rough edges
 of the absorber surface can then be treated as a time-dependent variation
 of the \emph{flat} absorber position. This means that the neutron loss
 mechanism in such a model is equivalent to the ionization of a
 particle, initially confined in a well with an oscillating wall.

 The time-dependent Schr\"{o}dinger equation for the neutron
 wave-function is:
 \begin{equation}\label{TdSchrod}
 i\frac{\partial \Phi(t,z)}{\partial
 t}=\left[-\frac{1}{2m}\frac{\partial^2}{\partial
 z^2}+mgz+V(z,H(t))\right]\Phi(t,z)
 \end{equation}

 The time-dependence appears here through the time-dependence of the
absorber position $H(t)$.

 The boundary conditions are:
 \begin{eqnarray}\label{Mir}
 \Phi(t,z=0)&=&0\\
 \Phi(t,z=\infty)&=&0
 \end{eqnarray}

 It would be convenient here to introduce time-dependent basis functions:
 \[\phi_n(t,z)=\psi_n(H(t),z) \exp(-\frac{i}{\hbar}\int_0^{t}
 \varepsilon_n(H(\tau)d\tau) \]

 where $\psi_n(H,z)$ and $\varepsilon_n(H)$ are \emph{complex}
eigenfunctions and eigenvalues of transversal Hamiltonian
(\ref{trans}) with fixed
 absorber position $H$.

 The total wave-function $\Phi(t,z)$ can be expanded in the set of functions:
 \begin{equation}\label{PhiTemp}
 \Phi(t,z)=\sum_n C_n(t)\psi_n(H(t),z) \exp(-i\int_0^{t}
 \varepsilon_n(H(\tau))d\tau)
 \end{equation}

 The equation system for the expansion factor $C_n(t)$ is:
 \begin{eqnarray}\label{Ct}
 \frac{dC_n(t)}{dt}&=&-\frac{dH}{dt}\sum_{k\neq
 n}C_k \alpha_{nk}\exp[-i\omega_{nk}(t)]\\
 \alpha_{nk}&=&-\alpha_{kn}\equiv
 \int_0^{\infty}\psi_n(H,z)\frac{\partial\psi_k(H,z)}{\partial H}
 \label{alphaNK}
 \end{eqnarray}

 where

 \[
 \omega_{nk}(t)=\int_0^{t}
 [\varepsilon_k(H(\tau))-\varepsilon_n(H(\tau))]d\tau
 \]
  Note that the derivation of equations (\ref{Ct}) and
 (\ref{alphaNK}) requires the biorthogonal condition (\ref{BiOrt}).

 The initial conditions $C_k(0)$ are determined by the overlapping of
 the incoming flux with the basis  functions:
 \[
 C_k(0)=\int_0^{\infty} \Phi(t=0,z) \psi_k(H,z) dz
 \]
  The solution to equation (\ref{Ct}) together with the above initial
  conditions enables us to obtain the wave-function $\Phi(t,z)$ at
$t=\tau^{pass}$
  and to calculate the measured flux:
  \[F=\int_0^{\infty} |\Phi(\tau^{pass},z)|^2 dz \]

 The equation system (\ref{Ct}) can be very much simplified under the
 following assumptions.

 First, let us suggest that the absorber position time-dependence is
 harmonic:
 \[ H(t)=H_0+b \sin(\omega t) \]

 where $b$ is the roughness amplitude and frequency $\omega=V/d$, with
 $d$ being the spatial period of roughness. It is known that in such
 cases it is only the states in the equation system (\ref{Ct}) obeying the
"resonance" condition:
 \begin{equation}
 \label{res} |\varepsilon_k-\varepsilon_n|\simeq \omega
 \end{equation}

 which are effectively coupled.
 As long as the transversal states have the widths  this resonance can
 not be exact. However we will restrict our treatment to only two
 coupled states.

 Second, we expect the roughness amplitude to be so small that the
 following approximation is valid:
 \[ \frac{d \psi_k(H(t),z)}{dt}\simeq b \cos(\omega
t)\frac{\partial\psi_k(H,z)}{\partial
 H}|_{H=H_0}\]

 Thirdly, we will consider that $ \omega \gg |\varepsilon_n|$, so that
 the low lying gravitational state $\psi_n$ is coupled with the very
 highly excited state with energy $\mathop{\rm Re}\varepsilon_k \gg
 MgH$. The gravitational potential can be neglected in comparison
 with such high energy; we are thus dealing with a "box-like" state.
 Its energy and width is given by (\ref{Ebox}) and (\ref{Gbox}). As
 the width of such excited states is much bigger than the
 width of the low lying gravitational state $\psi_n$ we can neglect
  the latter and suppose that neutrons in the low-lying
 gravitational state are \emph{elastically} reflected both from the
 mirror and the absorber. This results in the following boundary
 conditions for the gravitation state wave-function $\psi_n(H,z)$:
 \[\psi_n(H,z=0)=\psi_n(H,z=H)=0\]

 The eigenfunction $\psi_n(H,z)$ and the eigenvalue
 $\varepsilon_n(H)=\varepsilon_0 \lambda_n(H)$, determined by the
 above boundary condition are:

 \begin{eqnarray}\label{Pn}
 \psi_n(H,z)&\sim& \mathop{\rm Bi}(-\lambda_n(H))\mathop{\rm
 Ai}(z-\lambda_n(H))-\mathop{\rm Ai}(-\lambda_n(H))\mathop{\rm
 Bi}(z-\lambda_n(H))\\ \label{En}
  \mathop{\rm Ai}(H/l_0-\lambda_n(H))\mathop{\rm
 Bi}(-\lambda_n(H))&=&\mathop{\rm Ai}(-\lambda_n(H))\mathop{\rm
 Bi}(H/l_0-\lambda_n(H))
 \end{eqnarray}

 Finally we come to the equation system with only two coupled
 equations:

 \begin{equation}\label{cn2}
 \left\{
 \begin{array}{cll}
 \dot{C_0}(t)&=&-\frac{1}{2}b \omega C_1(t)\alpha(H)
 \exp[i(\omega-\omega_{01})t] \\
 \dot{C_1}(t)&=&\frac{1}{2}b \omega C_0(t)\alpha(H)
 \exp[-i(\omega-\omega_{01})t]
 \end{array}
 \right.
 \end{equation}
 with
 \[\alpha(H)=\int_0^{H}\psi_0(H,z)\frac{\partial\psi_1(H,z)}{\partial
 H}\] and $\omega_{01}=E_1-E_0$.

In the above expressions index $0$ labels the low lying
gravitational state, while index $1$
 labels the  excited fast decaying box-like state with \emph{complex} energy $E_1=\mathop{\rm Re}E_1-i\Gamma/2$.

 A very convenient expression \cite{ME} can be obtained for coupling matrix
 element $\alpha(H)$ (see Appendix B), namely:
 \begin{equation}\label{LambdLoc}
 \alpha(H)=\frac{\sqrt{\partial \lambda_0/\partial H \partial
 \lambda_1/\partial H}}{\lambda_0-\lambda_1}
 \end{equation}

 The benefit of such a simplified equation system is that it enables an
 analytical solution. Taking into account initial conditions
 $C_0(0)=1$ and $C_1(0)=0$ we get:
 \begin{eqnarray}
 C_0(t)&=&\exp\left(-\frac{\Gamma t}{4}\right) \left(\cos(\gamma
 t/2)+\frac{\Gamma}{2\gamma }\sin(\gamma t/2) \right) \label{cnsol}\\
 C_1(t)&=&-i \frac{\alpha(H)}{\gamma} \exp\left(\frac{\Gamma
 t}{4}\right)\sin(\gamma t/2) \label{cksol}
 \end{eqnarray}
  Here $\Gamma$ is the width of the "box-like" state and
  $\gamma=1/2 \sqrt{b^2 \omega^2 \alpha^2(H)-\Gamma^2}$.
(Note the exponential increase of $C_1(t)$. This does not yield in
nonphysical result, as far as in the expression for the
wave-function (\ref{PhiTemp}) $C_1(t)$ is multiplied by decaying
exponent $\exp(-iE_1 t)$. However, as we mentioned before,
$|C_0(t)|^2$ and $|C_1(t)|^2$ cannot be interpreted as probabilities
to find a system in certain quantum state).

  We are interested in the evolution of the gravitational state. Two
important limiting cases are:
 \begin{eqnarray}
 |C_0(t)|^2&\rightarrow&\exp(-\frac{\Omega^2 t}{4\Gamma})
 \mbox{, if } \Omega^2/\Gamma^2\ll 1 \label {cnsmall}\\
 |C_0(t)|^2&\rightarrow&\exp\left(-\frac{\Gamma
 t}{2}\right)\cos^2(\Omega t/2-\varphi)/\cos(\varphi) \mbox{, if
 }\Omega^2/\Gamma^2\gg 1 \label{cnlarge}
 \end{eqnarray}
 Here $\varphi=\arctan{(\Gamma/(2\gamma))}$ and $\Omega^2=b^2
 \omega^2 \alpha^2(H)$

 The quantity $\Omega$ plays the role of "transition frequency" between
  two states. It is proportional to the roughness amplitude $b$ and depends
on the averaged absorber position
  $H$
 via the coupling $\alpha(H)$. The coupling $\alpha(H)$ decays
 rapidly as soon as $H> H_n$, where $H_n$ is the classical turning point for
the low-lying gravitational state.

 When $\Gamma\gg \Omega$ the decay rate $\Gamma_n$ of the $n$-th
 gravitational state , according to (\ref{cnsmall}), is
 $\Gamma_n=\Omega^2/(4\Gamma)$.
Using the asymptotic expressions for
 $\alpha(H)$ (see Appendix B) one can get the following expression for
  the decay rate in case $H\gg H_n$:
  \begin{equation} \label{rateWKB}
 \Gamma_n=\varepsilon_0\sqrt{\frac{l_0}{H_n}}\frac{b^2}{8l_0|\mathop{\rm
 Im} a|}\sqrt{\frac{H-H_n}{l_0}}\exp \left[-\frac{4}{3}(
 (H-H_n)/l_0)^{3/2}\right]
 \end{equation}
  where $\mathop{\rm Im} a$ is the imaginary part
  of the scattering length of the neutrons on the \emph{flat absorber}
Fermi potential.

 The expression (\ref{rateWKB}) should be compared with the analogous
 formula for flat absorbers (\ref{Gfl}). One can see that
 \[
 a_{\rm eff}=\frac{b^2}{16|\mathop{\rm Im} a|}
 \]
 plays the role of the effective scattering length of the rough surface absorber,
 which is proportional to the \emph{square} of the roughness
 amplitude.


  The time
  \[ \tau^{abs}_n=\frac{8 l_0|\mathop{\rm
 Im} a|}{b^2\varepsilon_0}\sqrt{\frac{l_0}{H_n}}\] plays the role of
 the characteristic absorption time in our problem.

 It should be noted that the above results are true for $H>H_n$ and "weak"
coupling.
 When the absorber position $H<H_n$ the coupling is large and another
 limiting case applies, namely $\Omega\gg\Gamma$. In such cases
(\ref{cnlarge}) the gravitational state decay within the lifetime:
 \[\tau=2/\Gamma\]
 which is small compared to the passage time through the wave-guide.
 \begin{figure}
  \centering
 \includegraphics[width=75mm]{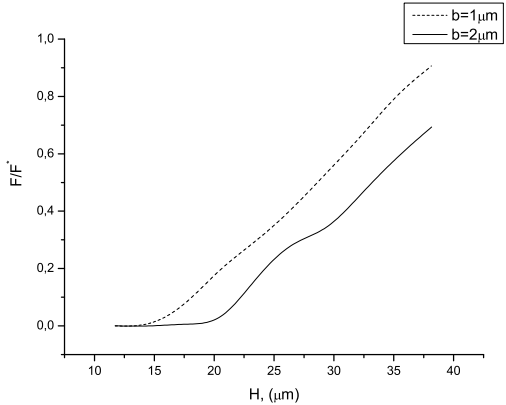}
  \caption{The relative neutron flux as a function of the slit height in
the time-dependent model. $F^*$ is the flux calculated for absorber
position $H=40$ $\mu$m,
  $b=1$ $\mu m$, $|\mathop{\rm Im} a|= 0.1$ $\mu m$ and $\tau^{pass}=0.02$
s. }\label{TimeDep}
 \end{figure}

  On Fig.\ref{TimeDep} we plot the results of numerical calculations for the
 measured neutron flux within the time-dependent model for 2 values
 of roughness amplitude $b=1$ $\mu m$ and $b=2$ $\mu m$ and
 $|\mathop{\rm Im} a|=0.1$ $\mu m$. Better resolution of the
 quantum "steps" appears with an increase of the roughness amplitude.

In the above simple "two-state" model several potentially important
effects are not taken into account, in particular the "non-resonant"
transitions between different gravitational states. However this
model enables understanding of fast irregularities ("steps") in the
transmitted neutron flux as a function of absorber position $H$ and
naturally explains them in terms of gravitational states of
neutrons. The model also establishes the dependence of the
wave-guide absorbing properties on roughness amplitude.

 \subsection{Resolution of gravitational states}
 \subsubsection{Constraints on resolution}

 Based on the results of the previous sections we can analyze the
 conditions for the best resolution of gravitational states. The
 presented numerical calculations show that an increase in absorber
 efficiency (e.g. by increasing roughness amplitude) results in
 a shifting of the positions of "the quantum steps" in the neutron flux by
 the value $\Delta_n$ and enhancing their resolution $\delta_n$. To
 perform a qualitative analysis we will accept that the "step-like"
 increase in the measured neutron flux, corresponding to the
 "appearance" of the new state, starts to be seen when the widths of
 this state are:
 \[\Gamma_n(H)\tau^{pass}=e\]
 We will also accept that such a "step-like" increase saturates when
 \[\Gamma_n(H)\tau^{pass}=1/e\]

  From expression (\ref{rateWKB}) one obtains the following estimate
 for the shift $\Delta_n$ and the uncertainty $\delta_n$ of the
 $n$-th step in the extreme limit $\ln (
 \tau^{pass}/\tau^{abs}_n)\gg 1$:
 \begin{eqnarray}\label{Delt}
 \Delta_n &\simeq& l_0\left(\frac{3}{4}\right)^{2/3} \left[ \ln (
 \tau^{pass}/\tau^{abs}_n\sqrt {\Delta_0/H_n}) \right]^{2/3} \\
 \delta_n&\simeq& \frac{2l_0}{3}\left(\frac{3}{4}\right)^{2/3}\left[
 \ln ( \tau^{pass}/\tau^{abs}_n\sqrt {\Delta_0/H_n}) \right]^{-1/3}
 \end{eqnarray}
  where
  \[
  \Delta_0=l_0\left(\frac{3}{4}\right)^{2/3} \left[ \ln (
 \tau^{pass}/\tau^{abs}_n)\right]^{2/3}
 \]

 The uncertainty $\delta_n$ decreases as $\ln ^{-1/3}(
 \tau^{pass}/\tau^{abs}_n)$ with an increase in
 $\tau^{pass}/\tau^{abs}_n$.
  The resolution of the $n$-th
 state is possible if the uncertainty in the step position
 $\delta_n$ is much less than the distance between neighboring steps
 $H_{n+1}+\Delta_{n+1}-H_n-\Delta_n$. For highly excited states we
 can use the WKB expression (\ref{WKB}) for the classical turning point:
 \[
 H_n=l_0\left(\frac{3\pi}{4}(2n-1/2)\right)^{2/3}
 \]
 to find the \emph{universal} limit on the number of states that can
 be resolved if $\tau^{pass}/\tau^{abs}_n\gg 1$:
 \[
 \left[ \ln ( \tau^{pass}/\tau^{abs}_n) \right]^{-1/3}\ll
 \frac{3}{2}( 2\pi)^{2/3}\left[(n+3/4)^{2/3}-(n-1/4)^{2/3}\right]
 \]

 This estimation shows that the resolution of states very slowly
 increases with an increase in passage time or in the efficiency of the
absorber
 in the limit $\tau^{pass}/\tau^{abs}_n\gg 1$, namely $n \sim \ln (
 \tau^{pass}/\tau^{abs}_n)$. This law is the consequence of the
 linear dependence of the gravitational potential on $z$. Indeed due to
 the linearity of the gravitational potential, the level spacing
 decreases with $n$ like $n^{-1/3}$, until the neighboring states'
 contribution to the flux starts to overlap. In particular for the
 value of $\tau^{pass}/\tau^{abs}_n=100$ the number of states that
 can be resolved is around $5$.

 The resolution of quantum states could be improved, if the initial
 population of one or several of such states is artificially reduced.
 In this case the neighboring state would be exposed. We have studied the
scenario in which the bottom mirror has a specially designed
 "step" \cite{Netal00,AW}.

 \subsubsection {Repopulation of states}

 If two bottom mirrors are shifted relative to each other by a
  $\Delta$ of a few $\mu m$ in height, there is an additional
 boundary at the step position $x=L_{0}$ that will change the
 population of the eigenstates. We give here just a brief
 description (for the details see \cite{AW}).

 \begin{equation}
\Psi_I\big|_{x=L_0}=\Psi_{II}\big|_{x=L_0}\;\;\wedge\;\;\left.\frac{\partial}{\partial x}\Psi_I\right|_{x=L_0}=\left.\frac{\partial}{\partial x}\Psi_{II}\right|_{x=L_0}\;\;\forall z\in\left[0,H\right]\nonumber\\
\label{match}
\end{equation}

 Due to the presence of the shift $\Delta$ in the bottom mirrors'
 position, the gravitational states are repopulated. If this step is
 treated as a "sudden change" in the potential, the matching at
 the boundary $x=L_0$ results in the following repopulation
 coefficients:
 \begin{equation}\label{Crep}
 C_{jm}=\exp\left(-\Gamma_j\tau_0/2)\right)\int_0^H
 \varphi_j(z)\varphi_m(z+\Delta) dz
 \end{equation}
 Here $\tau_0=L_0/V$, $\varphi_j(z)$ is the gravitational state in
 the presence of the absorber, positioned at height $H$ above the
 first mirror. Again, note the usage of the biorthogonality
 condition.

 The expression for the neutron flux at the detector position is now
 modified as follows:
 \begin{equation}\label{FlRep}
 F(L)\simeq \sum_{j,m}
 |C_{jm}|^2\exp\left(-(\Gamma_j-\Gamma_m)\tau_0-\Gamma_m
 \tau^{pass}\right)
 \end{equation}
 We neglect here the interference terms, assuming wide longitudinal
 velocities distribution.


 To illustrate the effect of repopulation, consider a simplified system
 consisting of just two mirrors without any absorber. The orthonormal
 system of eigenfunctions of the vertical motion in this case is just
 given by the standard bound state Airy function. Now imagine the
 second mirror shifted downwards compared to the first one
 by an amount equal to the height of the first node of the $2^{nd}$
 eigenstate wavefunction
 \[(\lambda_2-\lambda_1)\cdot l_0\approx 1.56\cdot l_0\approx 9.15\,\mu m\]
 It is clear that the $2^{nd}$ eigenstate above the $2^{nd}$
 mirror exactly matches the ground state wave function above the
 $1^{st}$ mirror from its edge on, while the ground state wave
 function of the $2^{nd}$ mirror overlaps only with the exponentially
 decaying tail of the ground state of the $1^{st}$ mirror. This implies
 immediately that the new ground state above the $2^{nd}$ mirror will
 be suppressed with respect to the $2^{nd}$ eigenstate above this
 mirror. The repopulation coefficients for the transition to the
 $2^{nd}$ mirror, normalized to the initial population of the ground
 state above the $1^{st}$ mirror, are given in Table\ref{Table2}.


 \begin{table}
 \centering
 \begin{tabular}{|c|l|}
  \hline
  $n$ & $(\left\langle\psi_n\right|\left.\psi_1\right\rangle)^2$\\
  \hline
  1 & 0.162 \\
  2 & 0.765 \\
  3 & 0.037 \\
  4 & 0.019 \\
  5 & 0.009 \\
  6 & 0.005 \\
  7 & 0.002\\
  \hline
 \end{tabular}
 \caption{Normalized repopulation coefficients after transition from
 a single ground state across a mirror shift of $\approx 9.15\,\mu
 m$} \label{Table2}
 \end{table}

\begin{figure}[htbp]
 \begin{center}
 \epsfig{file=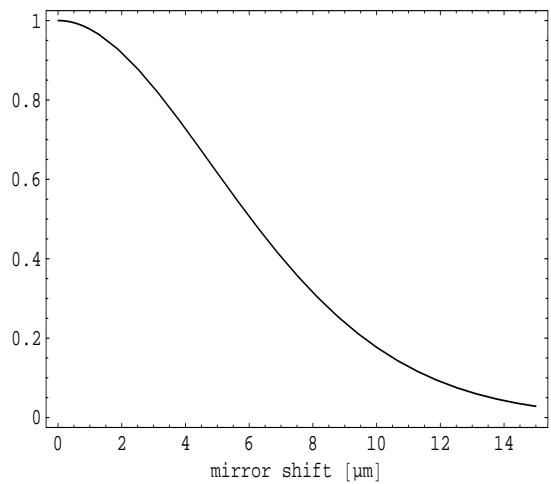,height=2.5in,width=3.0in}
 \caption{Repopulation coefficient of the ground state above the
 $2^{nd}$ mirror as a function of the relative shift of the two
 mirrors in $\mu m$}\label{repop}
 \end{center}
 \end{figure}

 Fig.\ref{repop} contains a plot of the population of the new ground
 state above the $2^{nd}$ as a function of the relative shift of the
 two mirrors.

\begin{figure}
  \centering
 \includegraphics[width=75mm]{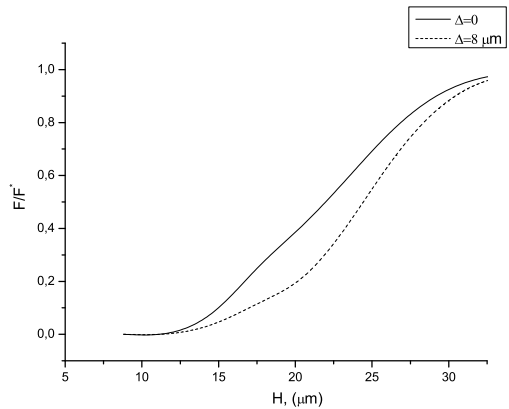}
  \caption{The relative neutron flux as a function of slit height for two
values of the bottom mirror shift. $F^*$ is the flux calculated at
absorber position $H=35$ $\mu m$ and diffuseness $\rho=1$ $\mu
m$.}\label{mirr}
 \end{figure}

 In Fig.\ref{mirr} the neutron flux for mirror shift $\Delta=8$ $\mu
 m$ is compared with the neutron flux without any shift in the bottom
 mirror position. Due to the depopulation of the ground state, the changes
 in the flux slope corresponding to the gravitational states can more
easily be
 seen.

 \section{Conclusions}

 We have analyzed the problem of the passage of ultra-cold neutrons through
an absorbing wave-guide in the presence of the Earth's gravitational
 field, both qualitatively and numerically. We have shown that the set of
existing experimental results \cite{Netal1,Netal2,Netal3} exhibits
clear evidence for the quantum motion of neutrons in the
gravitational field.

  We developed the formalism describing the loss mechanism of ultra-cold neutrons in the
   wave-guide with absorption. The essential role of the  quantum
   reflection phenomenon
 for the loss of ultra-cold neutrons was established. The concept of
 quantum reflection enables universal description of  different kind
 of absorbers in terms of effective complex scattering length $a$. The
 efficiency of absorption of ultra-cold neutrons in the presence of
 the
 gravitational field of Earth is determined by the ratio of such a scattering length to the
 characteristic gravitational wave-length $a/l_0$.

We studied the particular case of absorbers with rough
 surface.   It was established that in the latter case the main loss
 mechanism is due to the non-specular reflection of neutrons from the
 rough edges of the absorber. Absorber efficiency turns out to be
 proportional to the square of its roughness amplitude, if this
 amplitude is small compared to the characteristic gravitational
 wave-length $l_0$.

 We calculated the neutron flux through the wave-guide in
 the case of zero gravity (mirror and absorber arranged parallel
 to the gravitational field). For large slit
heights, the dependence of such a flux on the slit height H exhibits
a power law. Its  exponent depends on the absorber efficiency.
 These
 calculations are important for independent measurement of
 absorber/mirror properties.

 We argue the possibility of using the "inverse geometry"
 experiment for measuring the lifetime of neutrons bouncing on an
 absorbing surface. The neutron lifetime was found to be
 $\tau^{abs}=1/(2mg|\mathop{\rm Im}a|)$. It was determined by the
 gravitational force $mg$, acting on the neutron and imaginary part
 of the scattering length $|\mathop{\rm Im}a|$ of the absorbing
 surface Fermi potential. This experiment shows unambiguously the
 role of gravitation on the lifetime of ultra-cold neutrons.


 The theory developed in this paper allows  to analyze the resolution of
 the gravitational spectrometer  and to compare the efficiency of
different kinds of absorbers/scatterers. We show that the
spectrometer resolution is severely limited by a fundamental reason:
finite  penetrability of the gravitational barrier between the
classically allowed region and the scatterer height. The resolution
can be improved by a  significant increase in the time of storage of
neutrons in quantum states, and/or by  improvement of the efficiency
of the absorber/scatterer. The efficiency of best
absorbers/scatterers used in actual experiments was defined mainly
by the shape of their rough surface so that the efficiency is
approximately proportional to the square of the roughness amplitude
(when the roughness amplitude is smaller than the characteristic
scale of the gravitationally bound quantum states $l_0$ ). Further
increase of the roughness does not improve the efficiency; however
strict theoretical description of the case of a large amplitude
roughness is not covered by the present analysis.
 Another way of increasing the resolution could be through the
 selective depopulation of certain gravitational states, for instance
 by applying a bottom mirror with a "step".

 The results obtained are rather general in character and can be
 applied to different physical problems, involving the transmission of
 quantum particles through absorbing wave-guides. The development
 of the theoretical considerations presented would include the
 incorporation of large roughness amplitudes comparable to  or larger than
 the characteristic gravitational length  $l_0 \sim 6$
 $\mu$m; as well as the studies of  long storage time case, when
 decay of neutron quasi-bound gravitational states differs from
  the exponential law. These are necessary if the highest resolution is to be achieved
for the
 method considered.

\section{Acknowledgement}

The present work was supported by the INTAS grant 99-705 and by the
German Federal Ministry for Research and Education under contract
06HD153I. We are sincerely grateful to all the members of our
collaboration and those who have shown interest in this work and
stimulated its development.

 \section{Appendix A}

 Here we derive the equation for the energies of neutrons localized
 between an ideal mirror and absorber in the presence of a gravitational
field. We assume that the absorber Fermi potential has a diffuse
radius much smaller than the characteristic gravitational
wave-length $\rho\ll l_0$. In the region where the absorber
 potential can be fully neglected $0\leq z\ll H-\rho$ the wave-function is
the superposition of Airy functions:
 \[\psi_b(z)\sim\mathop{\rm Ai}(z/l_0-\lambda_n)-S \mathop{\rm
Bi}(z/l_0-\lambda_n)\]
 The zero boundary condition on the mirror gives:
 \begin{equation}\label{S1}
 S=\frac{\mathop{\rm Ai}(-\lambda_n)}{\mathop{\rm
 Bi}(-\lambda_n)}
 \end{equation}

 The neutron wave-function inside absorber $z> H-\rho$ is determined
 by the absorber Fermi potential, which is much stronger than the
 gravitational potential. In the range of distances $H-l_0\ll z\ll
 H-\rho$ such a wave-function is weakly perturbed by gravitation and
 can be written as:
 \[\psi_a(z)\sim 1+\frac{H-z}{\tilde{a}}\]
 where ${\tilde{a}}=a-H$, with $a$ being the complex scattering
 length on the absorber Fermi potential. One can see that $\tilde{a}$
 plays the role of the "scattering length of the diffuse tail" of the
 Fermi potential.

 Now we match the wave-functions $\psi_b(z)$ and $\psi_a(z)$ and
 their derivatives in the region $H-l_0\ll z\ll H-\rho$. For this we use
the Tailor expansion of $\psi_b(z)$ in the vicinity
 of $H$:
 \[\psi_b(z)\sim \mathop{\rm Ai}(H/l_0-\lambda_n)-S \mathop{\rm
 Bi}(H/l_0-\lambda_n)+\left(\mathop{\rm Ai'}(H/l_0-\lambda_n)-S
 \mathop{\rm Bi'}(H/l_0-\lambda_n)\right)(z-H)\]

 The matching condition gives:
 \begin{equation}\label{S2}
 S=\frac{\mathop{\rm
 Ai}(H/l_0-\lambda_n)-\tilde{a}/l_0\mathop{\rm
 Ai'}(H/l_0-\lambda_n)}{\mathop{\rm
 Bi}(H/l_0-\lambda_n)-\tilde{a}/l_0\mathop{\rm
  Bi'}(H/l_0-\lambda_n)}
 \end{equation}

 Putting together (\ref{S1}) and (\ref{S2}) we finally get the
 equation for the eigenvalues $\lambda_n$:
 \begin{equation}
 \frac{\mathop{\rm Ai}(-\lambda_n)}{\mathop{\rm
 Bi}(-\lambda_n)}=\frac{\mathop{\rm
 Ai}(H/l_0-\lambda_n)-\tilde{a}/l_0\mathop{\rm
 Ai'}(H/l_0-\lambda_n)}{\mathop{\rm
 Bi}(H/l_0-\lambda_n)-\tilde{a}/l_0\mathop{\rm
  Bi'}(H/l_0-\lambda_n)}
 \end{equation}

 The equation for the eigenvalues in an inverse geometry
 experiment can be obtained in a similar way. The wave-function
 outside the absorber $\psi_b(z)$ now vanishes at the mirror position
 $H$, which gives for $S_{inv}$:
 \begin{equation}\label{Sinv}
 S_{inv}=\frac{\mathop{\rm Ai}(H/l_0-\lambda_n)}{\mathop{\rm
 Bi}(H/l_0-\lambda_n)}
 \end{equation}

 The wave-function $\psi_a(z)$ of the neutron inside the absorber at
 the asymptotic distances $z\gg \rho$ is:
 \[\psi_a(z)\sim1-z/a\]
 The matching of $\psi_a(z)$ and $\psi_b(z)$ at distances $l_0\gg
 z\gg \rho$ together with (\ref{Sinv}) results in the following
 equation for $\lambda_n$:
 \begin{equation} \label{LambdInv}
 a\left[\mathop{\rm Ai}(H/l_0-\lambda_n)\mathop{\rm
  Bi'}(-\lambda_n)-\mathop{\rm Ai'}(-\lambda_n)\mathop{\rm
  Bi}(H/l_0-\lambda_n)\right]=\mathop{\rm Ai}(-\lambda_n)\mathop{\rm
  Bi}(H/l_0-\lambda_n)-\mathop{\rm Bi}(-\lambda_n)\mathop{\rm
  Ai}(H/l_0-\lambda_n)
 \end{equation}

 Note that the derivation of the above equations is based
 on the fact that $\rho\ll l_0$, so that the wave-function
 $\psi_a(z)$ is weakly perturbed by the gravitational field in the
 asymptotic region $\rho\ll z\ll l_0$.

 \section{ Appendix B}

 In this Appendix we derive the useful relation between the nonadiabatic
 coupling matrix element $\langle\varphi_j|\frac{\partial
 \varphi_i}{\partial H}\rangle $ and the energies of corresponding
 states $i$ and $j$.

 We will study the one dimensional Schr\"{o}dinger equation:
 \begin{equation}\hat{\textbf{H}}|\varphi_i\rangle=E_i|\varphi_i\rangle
 \label{Eq}
 \end{equation}
 The eigenfunctions $\varphi_j(x)$ and $\varphi_i(x)$
  obey the following boundary condition:
 \begin{eqnarray}
 \varphi_i(x=0)&=&0\label{zeroboundary} \\
 \varphi_i(x=H)&=&0 \label{Hboundary}
 \end{eqnarray}

 Here the varying parameter $H$ is a boundary. Hereafter we assume
 that the Hamiltonian itself is independent of $H$, while eigenfunctions
 $\varphi_i(x,H)$ and energy eigenvalues $E_i(H)$ depend on $H$
 through the boundary condition (\ref{Hboundary}) only.

 Applying $\partial/\partial H$ to both sides of (\ref{Eq})we get:
 \begin{equation}\label{EQder}
 \hat{\textbf{H}}\frac{\partial \varphi_i}{\partial H}=\frac{\partial
 E_i}{\partial H} \varphi_i+E_i\frac{\partial \varphi_i}{\partial H}
 \end{equation}

 Integrating the left side of (\ref{EQder}) with $\varphi_j(x,H)$ and taking into account
 boundary conditions (\ref{zeroboundary}) and (\ref{Hboundary}) we get:
 \[\langle \varphi_j|\hat{\textbf{H}}|\frac{\partial \varphi_i}{\partial
H}\rangle=
 -\frac{d\varphi_j(x)}{dx}\frac{\partial \varphi_i}{\partial
 H}\lfloor_{x=H}+ \langle \frac{\partial \varphi_i}{\partial
 H}|\hat{\textbf{H}}|\varphi_j\rangle\]

Note, that:
\[\langle \frac{\partial \varphi_i}{\partial
 H}|\hat{\textbf{H}}|\varphi_j\rangle=E_j\langle \frac{\partial \varphi_i}{\partial
 H}|\varphi_j\rangle
 \]
Combining the above results we get for the matrix element of
interest:
 \begin{equation}\label{matrel}
 \langle\varphi_j|\frac{\partial \varphi_i}{\partial
 H}\rangle=\frac{\frac{d\varphi_j(x)}{dx}\frac{\partial
 \varphi_i(x)}{\partial H}\lfloor_{x=H}+\frac{\partial E_i}{\partial
 H}\delta_{ij}}{E_j-E_i}
 \end{equation}

 Now let us use the following relation
 \[\frac{\partial\langle \varphi_i|\varphi_i\rangle}{\partial
 H}=2\langle \partial \varphi_i/\partial H|\varphi_i\rangle=0\]

 From (\ref{matrel}) we get in case $i=j$:

 \begin{equation}\label{diag}
 \frac{d\varphi_i(x)}{dx}\frac{\partial \varphi_i(x)}{\partial
 H}\lfloor_{x=H}=-\frac{\partial E_i}{\partial H}
 \end{equation}

 It is clear that the expression (\ref{matrel}) can be expressed as:
 \[
 \langle\varphi_j|\frac{\partial \varphi_i}{\partial
 H}\rangle=\frac{t_i t_j}{E_j-E_i}
 \]
  From (\ref{diag}) we finally get:
 \begin{equation}
 \langle\varphi_j|\frac{\partial \varphi_i}{\partial
 H}\rangle=\frac{\sqrt{\partial E_i/\partial H \partial E_j/\partial
 H}-\partial E_i/\partial H \delta_{ij}}{E_j-E_i}
 \end{equation}

 Applying the above result to the coupling matrix element in the
 time-dependent model (\ref{Ct}) we get:
 \begin{equation}
 \alpha(H)=\frac{\sqrt{\partial \lambda_n/\partial H \partial
 \lambda^{*}/\partial H}}{\lambda_n-\lambda^{*}}
 \end{equation}

 Here $\lambda_n$ is the eigenvalue of the low-lying gravitational state,
 while $\lambda^{*}$ is the eigenvalue of the highly excited "box-like"
 state. This expression is much more convenient for practical
 applications than the integral in the definition of the coupling
 matrix element. In particular, it can be used to obtain the
 asymptotic expressions for the width (\ref{rateWKB}) of a given
 gravitational state $n$ if $H\gg H_n$.

 To obtain such an expression we first find the eigenvalue
 derivative $\partial \lambda_n/\partial H$ from the equation
 (\ref{En}):
 \begin{equation}\label{deriv}
 \frac{\partial \lambda_n}{\partial H}=\frac{1}{l_0}\frac{\mathop{\rm
 Ai'}(H/l_0-\lambda_n)\mathop{\rm
  Bi}(-\lambda_n)-\mathop{\rm Bi'}(H/l_0-\lambda_n)\mathop{\rm
  Ai}(-\lambda_n)}{\mathop{\rm
 Ai'}(H/l_0-\lambda_n)\mathop{\rm
  Bi}(-\lambda_n)-\mathop{\rm Bi'}(H/l_0-\lambda_n)\mathop{\rm
  Ai}(-\lambda_n)+\mathop{\rm
 Ai}(H/l_0-\lambda_n)\mathop{\rm
  Bi'}(-\lambda_n)-\mathop{\rm Bi}(H/l_0-\lambda_n)\mathop{\rm
  Ai'}(-\lambda_n)}
 \end{equation}

 Taking into account the expression (\ref{En}) and asymptotic
 properties of the Airy function of large argument $H/l_0\gg
 \lambda_n$ we get:
 \[\frac{\partial \lambda_n}{\partial H}\approx
 -\frac{1}{l_0}\sqrt{\frac{H-H_n}{H_n}}\exp\left[-4/3(H/l_0-\lambda_n)^{3/2}
\right]
 \]

 For the energy $E^{*}$ of the highly excited "box-like" state, we
 can use expression (\ref{Ebox}), from which we get:
 \[\frac{\partial E^{*}}{\partial H}=-2\frac{E^{*}}{H} \]

 For the square of the coupling matrix element $\alpha^2(H)$ in case
 of large $H\gg H_n$ we get so far:
\[
\alpha^2(H)=2\frac{E^{*}}{H(E^{*}-E_n)^2}\frac{\varepsilon_0}{l_0}\sqrt{\frac{H-H_n}{H_n}}\exp\left[-4/3(H/l_0-\lambda_n)^{3/2}\right]
\]
  Taking into account the expression for the width $\Gamma^{*}$ of the
 "box-like" state (\ref{Gbox}) and substituting the above results
into the expression for the width of gravitational state
(\ref{cnsmall}):
\[\Gamma_n=b^2\omega^2\alpha^2(H)/(4\Gamma^{*})\]
 we finally come to the expression:
\[
\Gamma_n=\varepsilon_0\sqrt{\frac{l_0}{H_n}}\frac{b^2}{8l_0|\mathop{\rm
Im} a|}\sqrt{\frac{H-H_n}{l_0}}\exp \left[-\frac{4}{3}(
(H-H_n)/l_0)^{3/2}\right]
\]


\begin{references}
 \bibitem{Gold} Goldman, I. I. and Krivchenkov, V. D., Problems in Quantum
Mechanics. (London: Pergamon Press, 1961).
 \bibitem{Haar} D. ter Haar, Selected Problems in Quantum Mechanics (Academic,
 New York, 1964).
 \bibitem{Flugge} S. Flugge, Practical Quantum Mechanics I
 (Berlin,Springer, 1974)
 \bibitem{Langhoff}Langhoff, P. W. Schrodinger Particle in a Gravitational
Well.
 Am. J. Phys. 39 (1971), 954–957.
 \bibitem{Gibbs} Gibbs, R. L. The Quantum Bouncer. Am. J. Phys. 43 (1975),
25–28.
 \bibitem{Sakurai} J.J.Sakurai, Modern
 Quantum Mechanics (Benjamin/Cummings, Menlo Park, 1985), 26.
  \bibitem{Luschikov1} V.I.Luschikov (1977), Physics Today 30(6): 42.
 \bibitem{Luschikov2} V.I.Luschikov, A.I.Frank (1978). JETP Lett. 28(9): 559.
 \bibitem{Shull} C.G. Shull et al., Phys. Rev. \textbf{153} (1967)
 1415.
 \bibitem{Gahler} R. Gahler et.al. Phys.Rev. \textbf{D25} (1982) 2887.
 \bibitem{Bau}  J.
Baumann et al., Phys.Rev. \textbf{D37} (1988) 3107.


 \bibitem{Colella} R.Colella, A.W.Overhauser, S.A.Werner (1975). Phys. Rev.
Lett.
 34: 1472.
 \bibitem{Rauch} H.Rauch, H.Lemmel, M.Baron, R.Loidl
 (2002). "Measurement of a confinement induced neutron phase." Nature
 417 (6889): 630-632
 \bibitem{Stau} J.L. Staudenmann et al.,Phys. Rev. \textbf{A21} 1419
 (1980).
 \bibitem{Lit}  K.C. Litrell et al., Phys.Rev. \textbf{A56} 1767
(1997).
\bibitem{Zouw} X.Y. Zouw et al., Phys. Rev. Lett.
\textbf{69} 3041 (2000).


 \bibitem{Netal00} V.V. Nesvizhevsky et.al.,Nucl.Instr. Meth. \textbf{A440},
 754(2000)
 \bibitem{Netal1} V.V. Nesvizhevsky et.al., Nature \textbf{415}, 297
 (2002)
 \bibitem{Netal2} V.V. Nesvizhevsky et.al.,Phys.Rev. \textbf{D67}, 102002-1
 (2003)
 \bibitem{Netal3} V.V. Nesvizhevsky et.al., Eur. Phys. J. \textbf{C40}, 479
 (2005)
  \bibitem{Netal4}V.V.Nesvizhevsky, "Investigation of
 the neutron quantum states in the earth's gravitational field."
 Journal of Research of the National Institute of Standards and
 Technology 110(3), 263-267 (2005)
 \bibitem{Murayama}H.Murayama, G.G.Raffelt, C.Hagmann, K. van Bibber, and
 L.J.Rosenberg (2002), in Review of Particle Physics, Phys. Rev. D
 66: 344.
 \bibitem{Bertolami} O.Bertolami, F.M.Nunes (2003).
 "Ultracold neutrons, quantum effects of gravity and the weak
 equivalence principle." Classical and Quantum Gravity 20(5): 61-66.
 \bibitem {Bert1} O. Bertolami et. al., Phys. Rev. \textbf{D72},
 025010 (2005)
 \bibitem{EQ1} R. Aldrovandi, P. B. Barros, and J. G. Pereira "The
Equivalence Principle Revisited"
 arXiv:gr-qc/0212034 (2002)
 \bibitem{EQ2} Andrzej Herdegen, Jaroslaw Wawrzycki ,Is Einstein's
equivalence principle valid for a quantum
 particle?,arXiv:gr-qc/0110021 (2003)
 \bibitem{EQ3} R. Onofrio and L. Viola Phys. Rev. \textbf{D 55}, 455
 (1997)
 \bibitem{Abele} H.Abele, S.Bae{\ss}ler,
 and A.Westphal "Quantum states of neutrons in the gravitational
 field and limits for non-Newtonian interaction in the range between
 1 micron and 10 micron." Lect.Notes Phys. 631: 355-366 (2003)
\bibitem{BO} E. Narevicius, P. Serra, N. Moiseyev Europhys. Lett.
\textbf{62} (6), 789 (2003)
 \bibitem{NP1} Nesvizhevsky V.V., Protasov K.V. "Constraints on
 non-Newtonian gravity from the experiment on neutron quantum states
 in the earth's gravitational field." Classical and Quantum Gravity
 21: 4557-4566,(2004)

 \bibitem{NP2} V.V.Nesvizhevsky, and K.V.Protasov "Constraints on
 non-Newtonian gravity from the experiment on neutron quantum states
 in the earth's gravitational field." Journal of Research of the
 National Institute of Standards and Technology 110(3):
 269-272,(2005).

 \bibitem{Mavromatos} N.E.Mavromatos "CPT Violation and Decoherence in
Quantum
 Gravity." gr-qc/0407005 (2004).
 \bibitem{Rob} R.W. Robinett Phys. Rep. 392 (2004) 1-119
 \bibitem{Techn1}V.V.Nesvizhevsky
 (2005). "Polished sapphire for ultracold neutron guides." submitted
 to Nuclear Instruments and Methods.
 \bibitem{Techn2} C.Plonka et al
 (2005). (unpublished)
 \bibitem {Apllications} V.V.Nesvizhevsky and
 K.V.Protasov, in Edited Book 'Progress in Quantum Gravity', Frank
 Columbus, Nova, 2005




 \bibitem{RS1}S.K.Sinha, E.B.Sirota, and S.Garoff "X-ray and neutron
 scattering from rough surfaces." Physical Review \textbf{B 38}(4),
 2297-2312 (1988).

 \bibitem{RS2} R.Pynn "Neutron scattering by rough surfaces at grazing
 incidence." Physical Review \textbf{B 45}(2), 602-614, (1992)

 \bibitem{RS3} J.A.Sanchez-Gil, V.Freilikher, A.A.Maradudin, I.V.Yurkevich
  "Reflection and transmission of waves in surface-disordered
 wave-guides." Physical Review \textbf{B 59}(8), 5915-5925 (1999).

 \bibitem{RS4} N.M.Makarov, and A.V.Moroz "Spectral theory of a
 surface-corrugated electron waveguide: The exact scattering-operator
 approach." Physical Review \textbf{B 60}(1), 258-269 (1999).

 \bibitem{RS5} A.E.Meyerovich, and A.Stepaniants "Quantized systems
 with randomly corrugated walls and interfaces." Physical Review
 \textbf{B 60}(12), 9129-9144 (1999).

 \bibitem{Sweden} J.Hansson, D.Olevik, C.Turk, and H.Wiklund "Comment on
 "Measurement of quantum states of neutrons in the Earth's
 gravitational field"." Physical Review \textbf{D 68} (10),
 108701-108703, (2003).

 \bibitem{Answer} V.V.Nesvizhevsky et al.
 "Reply to "Comment on "Measurement of quantum states of neutrons in
 the Earth's gravitational field"." Physics Review D 68:
 108702(1-3)(2003).

 \bibitem{Bowles} J.T.Bowles "Quantum effects of gravity." Nature
 415(6869): 267-268 (2002).

 \bibitem{Schw} B.Schwarzschild "Ultracold neutrons exhibit quantum
 states in the Earth's gravitational field." Physics Today 55(3), 20
 (2002).

 \bibitem{Nanopart} V.V.Nesvizhevsky "Quantum states of neutrons in the
 gravitational field and interaction of neutrons with nanoparticles."
 Uspekhi Fizicheskikh Nauk 46(1): 93-97 (2003) (in Russian).

 \bibitem{NesvUspekh} V.V.Nesvizhevsky "Investigation of quantum neutron
 states in the terrestrial gravitational field above a mirror."
 Uspekhi Fizicheskikh Nauk 47(5): 515-522 (2004)(In Russian).


 \bibitem{AbSt} M. Abramowitz and I.E. Stegun {\it Handbook of mathematical
 Functions} (Dover Publ., New York 1965)
 \bibitem{GKK} I.I. Goldman, V.D. Krivchenkov, V.I. Kogan, V.M.
 Galitscii, Problems in Quantum Mechanics (New York, Academic, 1960)


 \bibitem{AW} A. Westphal, Diploma Thesis, Univ. of Heidelberg, 2001,
 gr-qc/0208062 (2003)
 \bibitem{KPV}V.A. Karmanov, K.V. Protasov, A.Yu. Voronin, Eur. Phys.
J.\textbf{ A 8}, 429 (2000)
 \bibitem{QR} J.E. Lennard-Jones, Trans. Faraday Soc. \textbf{28},333
 (1932)
 \bibitem{QR1} R. C\^{o}ter, H. Friedrich and J. Trost, Phys. Rev.
 \textbf{A56},1781 (1997)
 \bibitem{LL} L.D. Landau, E.M. Lifshitz, Quantum Mechanics (Oxford,
 Pergamon, 1976)

\bibitem{ME} O.I. Kartavtsev, private communications; E.A. Soloviev
Uspekhi Fizicheskikh Nauk 157 (3),437 (1989) (in Russian)






 \end{references}
\end{document}